\documentclass{IEEEoj}
\def\OJlogo{\vspace{-4pt}\hskip-4pt\includegraphics[height=18pt]{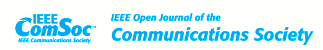}}
\usepackage{cite}
\usepackage{amsmath,amssymb,amsfonts}
\usepackage{algorithmic}
\usepackage{graphicx}
\usepackage{textcomp}
\usepackage[dvipsnames]{xcolor}
\usepackage{url}
\usepackage{tikz}
\usepackage{todonotes}
\usepackage{float}
\usepackage{subfigure}
\usepackage{siunitx}
\DeclareSIUnit{\gbps}{Gb/s}
\DeclareSIUnit{\gb}{Gb}
\usepackage{tabularx}
\usepackage{csquotes}
\usepackage{booktabs}
\usepackage{flushend}
\usepackage{lettrine}

\usepackage{scalerel}
\usepackage{tikz}
\usetikzlibrary{svg.path}
\definecolor{orcidlogocol}{HTML}{A6CE39}
\tikzset{
  orcidlogo/.pic={
    \fill[orcidlogocol] svg{M256,128c0,70.7-57.3,128-128,128C57.3,256,0,198.7,0,128C0,57.3,57.3,0,128,0C198.7,0,256,57.3,256,128z};
    \fill[white] svg{M86.3,186.2H70.9V79.1h15.4v48.4V186.2z}
                 svg{M108.9,79.1h41.6c39.6,0,57,28.3,57,53.6c0,27.5-21.5,53.6-56.8,53.6h-41.8V79.1z M124.3,172.4h24.5c34.9,0,42.9-26.5,42.9-39.7c0-21.5-13.7-39.7-43.7-39.7h-23.7V172.4z}
                 svg{M88.7,56.8c0,5.5-4.5,10.1-10.1,10.1c-5.6,0-10.1-4.6-10.1-10.1c0-5.6,4.5-10.1,10.1-10.1C84.2,46.7,88.7,51.3,88.7,56.8z};
  }
}
\newcommand\orcidicon[1]{\href{https://orcid.org/#1}{%
\mbox{\begin{tikzpicture}[yscale=-1,transform shape, scale=0.03] 
\pic{orcidlogo};
\end{tikzpicture}}}}

\usepackage[dvipsnames]{xcolor}

\usepackage{hyperref}
\hypersetup{
	colorlinks = true,
	allcolors=blue,
	urlcolor=black,
	linkcolor=black
}

\usepackage[nohyperlinks]{acronym}

\newcommand*\ballnumber[1]{\tikz[baseline=(char.base)]{
		\node[shape=circle,fill,inner sep=.5pt] (char) {\textcolor{white}{#1}};}}

\def\BibTeX{{\rm B\kern-.05em{\sc i\kern-.025em b}\kern-.08em
    T\kern-.1667em\lower.7ex\hbox{E}\kern-.125emX}}

\newcommand\citeN\cite

\newcommand\fig[1]{Figure~\ref{fig:#1}}

\newcommand\sect[1]{Section~\ref{sec:#1}}
\newcommand\equa[1]{Equation~\ref{eq:#1}}
\newcommand\tabl[1]{Table~\ref{tab:#1}}

\newcommand{\FI}[1]{\textcolor{black}{#1}}

\newcommand{\revieweri}[1]{\textcolor{black}{#1}}
\newcommand{\reviewerii}[1]{\textcolor{black}{#1}}

\newcommand{\figeps}[3][]{%
 \begin{figure}[htb!]
  \begin{center}
   \leavevmode
      \parbox[t]{#1}{%
        \resizebox{#1}{!}{\includegraphics{figures/#2}}
      }
   \caption{#3\vspace{-0.2cm}}
   \label{fig:#2}
  \end{center}
 \end{figure}
}

\newcommand{\foursubfigepsScale}[9]{
  \begin{figure}[htb!]
    \leavevmode
    \begin{center}
     \subfigure[#2]{
        \label{fig:#1}
        \parbox[t]{0.21\columnwidth}{%
            \resizebox{0.25\columnwidth}{!}{\includegraphics[width=0.25\columnwidth]{figures/#1}}
     \vspace{-1cm}
        }
     }
     \subfigure[#4]{
        \label{fig:#3}
        \parbox[t]{0.21\columnwidth}{%
            \resizebox{0.25\columnwidth}{!}{\includegraphics[width=0.25\columnwidth]{figures/#3}}
     \vspace{-1cm}
        }
     }
     \subfigure[#6]{
        \label{fig:#5}
        \parbox[t]{0.21\columnwidth}{%
            \resizebox{0.25\columnwidth}{!}{\includegraphics[width=0.25\columnwidth]{figures/#5}}
        }
     }
     \subfigure[#8]{
        \label{fig:#7}
        \parbox[t]{0.21\columnwidth}{%
            \resizebox{0.25\columnwidth}{!}{\includegraphics[width=0.25\columnwidth]{figures/#7}}
        }
     }
    \end{center}
     \vspace{-0.3cm}
    \caption{#9}
    \label{fig:#1_full}    
  \end{figure}
}

\newcommand{\twosubfigeps}[5]{
  \begin{figure}[htb]
    \leavevmode
    \begin{center}
     \subfigure[#2]{
        \label{fig:#1}
        \parbox[t]{0.47\textwidth}{%
            \resizebox{0.45\textwidth}{!}{\includegraphics{figures/#1}}
     \vspace{-1cm}
        }
     }
     \subfigure[#4]{
        \label{fig:#3}
        \parbox[t]{0.47\textwidth}{%
            \resizebox{0.45\textwidth}{!}{\includegraphics{figures/#3}}
     \vspace{-1cm}
        }
     }
    \end{center}
    \vspace{-0.2cm}
    \caption{#5}
    \label{fig:#1_full}
  \end{figure}
}

\newcommand{\twosubfigepsSideBySide}[5]{
  \begin{figure}[htb]
    \leavevmode
    \begin{center}
     \subfigure[#2]{
        \label{fig:#1}
        \parbox[t]{0.46\columnwidth}{%
            \resizebox{0.46\columnwidth}{!}{\includegraphics{figures/#1}}
     \vspace{-1cm}
        }
     }
     \subfigure[#4]{
        \label{fig:#3}
        \parbox[t]{0.46\columnwidth}{%
            \resizebox{0.46\columnwidth}{!}{\includegraphics{figures/#3}}
     \vspace{-1cm}
        }
     }
    \end{center}
    \vspace{0cm}
    \caption{#5}
    \label{fig:#1_full}
  \end{figure}
}

\newboolean{makevspace}
\newcommand{\cvspace}[1]{%
    \ifthenelse
        {\boolean{makevspace}}
        {\vspace{#1}}
        {}%
    }

\renewcommand{\thesection}{\Roman{section}}            

\begin{document}

\receiveddate{09 February 2025}
\reviseddate{07 March 2025}
\accepteddate{29 March 2025}
\publisheddate{02 April 2025}
\currentdate{23 May, 2025\\ \\ This is a consolidated version of the originally published article~\cite{IhMe25} and the official erratum~\cite{erratum}}
\doiinfo{OJCOMS.2025.3557082}

\jvol{6}
\pubyear{2025}

\title{MPLS Network Actions: Technological Overview and P4-Based Implementation on a High-Speed Switching ASIC}

\author{FABIAN IHLE$^{\orcidicon{0009-0005-3917-2402}}$ AND MICHAEL MENTH$^{\orcidicon{0000-0002-3216-1015}}$\IEEEmembership{(Senior Member, IEEE)}}
\affil{Chair~of~Communication~Networks, University~of~Tuebingen, 72076~Tuebingen, Germany}
\corresp{CORRESPONDING AUTHOR: M. Menth (e-mail: menth@uni-tuebingen.de).}
\authornote{This work was supported in part by the Deutsche Forschungsgemeinschaft (DFG) under Grant ME2727/3-1, and in part by the Open Access Publishing Fund of the University of Tübingen.
}
\markboth{MPLS NETWORK ACTIONS: TECHNOLOGICAL OVERVIEW AND P4-BASED IMPLEMENTATION}{IHLE and MENTH}

%


\begin{abstract}
In MPLS, packets are encapsulated with labels that add domain-specific forwarding information.
Special purpose labels were introduced to trigger special behavior in MPLS nodes but their number is limited.
Therefore, the IETF proposed the MPLS Network Actions (MNA) framework.
It extends MPLS with new features, some of which have already been defined to support relevant use cases.
This paper provides a comprehensive technological overview of MNA concepts and use cases.
It compares MNA to IPv6 extension headers (EHs) that serve a similar purpose, and argues that MNA can be better deployed than EHs.
It then presents P4-MNA, a first hardware implementation running at 400 Gb/s per port.
Scalability and performance of P4-MNA are evaluated, showing negligible impact on processing delay caused by network actions.
Moreover, the applicability of MNA is demonstrated by implementing the use cases of link-specific packet loss measurement using the alternate-marking-method (AMM) and bandwidth reservation using network slicing.
We identify header stacking constraints resulting from hardware resources and from the number of network actions that must be supported according to the MNA encoding.
They make an implementation for hardware that can only parse a few MPLS headers infeasible.
We propose to make the number of supported network actions a node parameter and signal this in the network.
Then, an upgrade to MNA is also feasible for hardware with fewer available resources.
We explain that for MNA with in-stack data (ISD), some header bits must remain unchanged during forwarding, and give an outlook on post-stack data (PSD).
\end{abstract}


\maketitle

\section{Introduction}
\IEEEPARstart{T}{he} Multiprotocol Label Switching (MPLS) protocol defined in RFC~3031~\cite{rfc3031} has become a predominant wide-area networking technology.
In MPLS, packets are equipped with labels adding domain-specific forwarding information. 
Packets are switched to intermediate nodes based on the assigned label.
Since the standardization in 2001, the protocol has continuously evolved.
In the last years, new applications have emerged that require labels to not only contain forwarding information but also information on how to process a packet.
Therefore, special purpose labels (SPLs) have been defined.
Examples of applications using SPLs include entropy labels which support \ac{ECMP}~\cite{RFC6790}, and labels for No Further Fast Reroute (NFFRR)~\cite{nffrr} which avoid the looping of packets during network failures. 
Further, labels can embed \ac{SFC} information\cite{rfc8595,HaSt22}, and data related to \ac{OAM}~\cite{mpls-oam,MaMu21}.

The IETF has proposed the \ac{MNA} framework~\cite{ietf-mpls-mna-fwk-05} which is currently in the process of standardization to facilitate extensions in the \acs{MPLS} protocol.
The \acs{MNA} framework provides an encoding for network actions and their data in the \acs{MPLS} label stack.
These network actions and their data can be either encoded as \ac{ISD} in the \acs{MPLS} stack or as \ac{PSD} after the \acs{MPLS} stack.
Since the \acs{MNA} framework serves as a foundation for future extensions of \acs{MPLS}, it must be efficiently supported by hardware.
\acp{EH} in IPv6 are a similar concept to the network actions introduced in the \acs{MNA} framework. 
However, various works \cite{NaGa22, CuSe24, GaNa22, rfc9098, HeVe17} have shown that IPv6 \acp{EH} are not widely adopted because they are not efficiently implementable on hardware.

The contribution of this paper is manifold.
We provide an introduction to the current state of the MNA framework and the first use cases identified by the working group.
We compare the concept of the MNA framework to IPv6 \ac{EH} and explore the feasibility for hardware implementations.
Then, we implement the \acs{MNA} framework based on the IETF MPLS working group (WG) proposal~\cite{ietf-mpls-mna-hdr-04} in P4 on the Intel Tofino™ 2 switching ASIC using \acs{ISD}.
Our implementation is the first and to date the only implementation of the MNA framework.
We evaluate the scalability and performance of the P4-MNA implementation.
Further, we implement and evaluate two example network actions, namely the \acf{AMM} for performance measurement, and bandwidth reservation using network slicing.
Based on our implementation experience, we propose to extend the signaling of hardware capabilities within an MNA-capable MPLS domain so that MPLS nodes with less capabilities can be integrated.
Finally, we identify challenges arising from using \acs{ISD}, propose solutions to address them in future work, \reviewerii{and summarize security considerations of the MNA framework}.

The rest of the paper is structured as follows. 
In \sect{background}, we provide a primer on the \acs{MNA} framework, including information on \acs{MPLS}, SR-MPLS, SPLs, and \acs{MNA}.
An overview of identified use cases by the IETF MPLS WG and their mechanisms is given in \sect{use_cases} and in \sect{related_work}, we compare IPv6 EH with MNA and review related work.
In \sect{p4}, we explain the concept of the programming language \acs{P4}.
The implementation of the \acs{MNA} framework on the Intel Tofino™ 2 is described in \sect{implementation}. 
In \sect{evaluation} we evaluate the P4-MNA implementation and the exemplary network actions.
In \sect{eval_signaling}, we describe constraints for header stacking and propose a signaling extension, and in \sect{challenges}, we identify challenges with mutable data.
\reviewerii{In \sect{security}, we describe security risks currently considered in the MNA framework.}
Finally, we conclude the paper in \sect{conclusion}.

\section{A Primer on MPLS Network Actions (MNA)}
\label{sec:background}
In this section, we give a brief overview of traditional MPLS networks including forwarding in MPLS and special-purpose labels.
Then, we explain the concept of the \acf{MNA} framework as being standardized by the IETF MPLS WG\cite{mpls_wg}. 
This includes the proposed \acs{MNA} header encoding, and the placement of network actions in the \acs{MPLS} label stack.
\subsection{Traditional MPLS Networks}
This section briefly summarizes the current state of the art for forwarding and extensions in MPLS networks.
\subsubsection{MPLS Forwarding}
Nodes in an MPLS network are called \acfp{LSR}. 
Initially, virtual circuits have been set up in MPLS along a path of \acp{LSR}.
These virtual circuits are called \acfp{LSP} and the virtual connection identifier is called a label.
The ingress and egress of an \acs{LSP} are called ingress \acf{LER} and egress \acs{LER}.
\acp{LSP} are established through signaling prior to communication.
Packets are forwarded by \acp{LSR} using label switching according to the entries of a forwarding table that contains information about incoming interface and label, and outgoing interface and label.
At the egress \ac{LER}, the MPLS label of the \acs{LSP} is popped and the packet is delivered to its destination using the underlying Layer 3 protocol.
An exemplary network using virtual circuits in MPLS is shown in \fig{pdfs/ihle1}.

\twosubfigeps{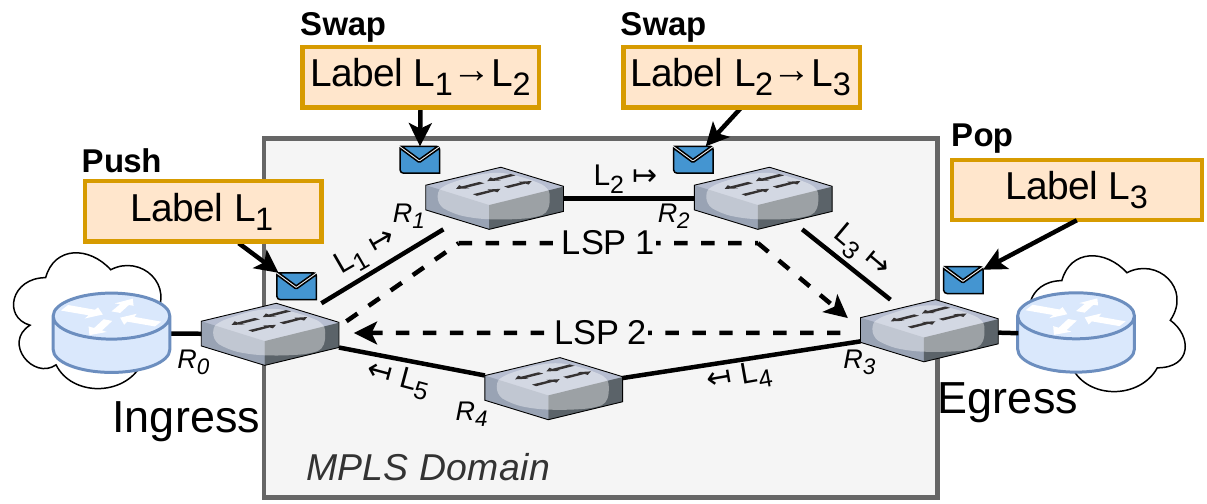}{\acp{LSP} are set up prior to communication and labels are switched according to \acs{LSP}-specific entries in the forwarding tables.}{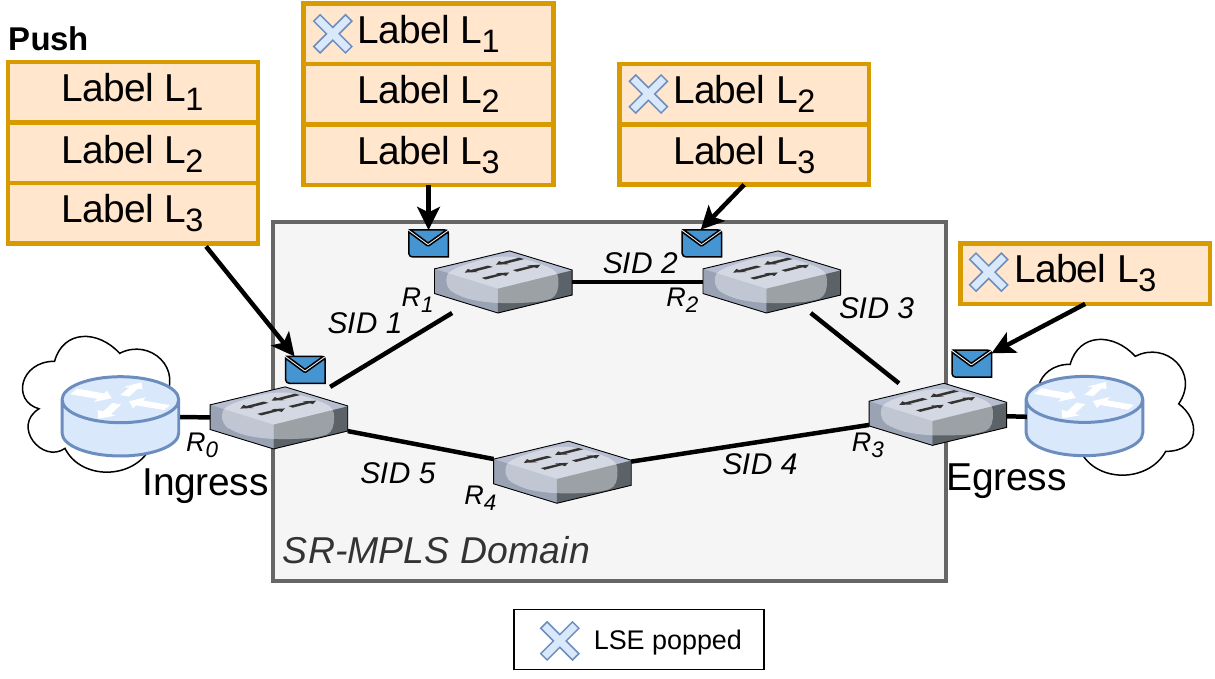}{With SR-MPLS, paths are encoded in packet headers as label stacks. During forwarding, \acp{LSR} pop the top-of-stack label.}{MPLS forwarding using virtual circuits and SR-MPLS.}

About 10 years ago, \ac{SR} has been introduced which turned MPLS into a connection-less and source-routed packet switching technology.
Labels are interpreted as segment identifiers (SIDs) which denote next intermediate nodes to where the packet is forwarded on the way to the egress \acs{LER}.
In \ac{SR}, there are two types of segments.
An adjacency segment describes a strict forwarding instruction over a specific link between two nodes while a prefix segment describes a loose forwarding instruction to a prefix over multiple hops~\cite{rfc8402}.
For an adjacency segment, a node pops the top-of-stack label and forwards the packet to the next segment~\cite{rfc8660}.
For a prefix segment, the top-of-stack label is not popped until the prefix is reached.
By stacking multiple labels, a source route is defined.
Thus, with \ac{SR}, the ingress \acs{LER} pushes a label stack and intermediate hops pop these labels.
There is no connection concept with SR-MPLS and therefore no signaling protocol is required to set up connections.
However, SIDs need to be signaled and label stacks need to be computed.
An exemplary network using \acs{SR}-MPLS with adjacency segments is shown in \fig{pdfs/ihle2}.
In both examples, penultimate hop popping is not applied, i.e., the egress node receives a label, pops it, and forwards the packet based on the underlying Layer 3 information.

The ingress \acs{LER} pushes one or more MPLS labels onto the MPLS stack of a packet.
An entry in this header stack is called a \acf{LSE}.
In \fig{pdfs/ihle3}, the MPLS header is visualized.
It is \SI{4}{\byte} large and consists of a 20 bit label, 3 bits traffic class (TC), a bottom-of-stack bit (S), and a 8 bit time-to-live (TTL).

\figeps[\columnwidth]{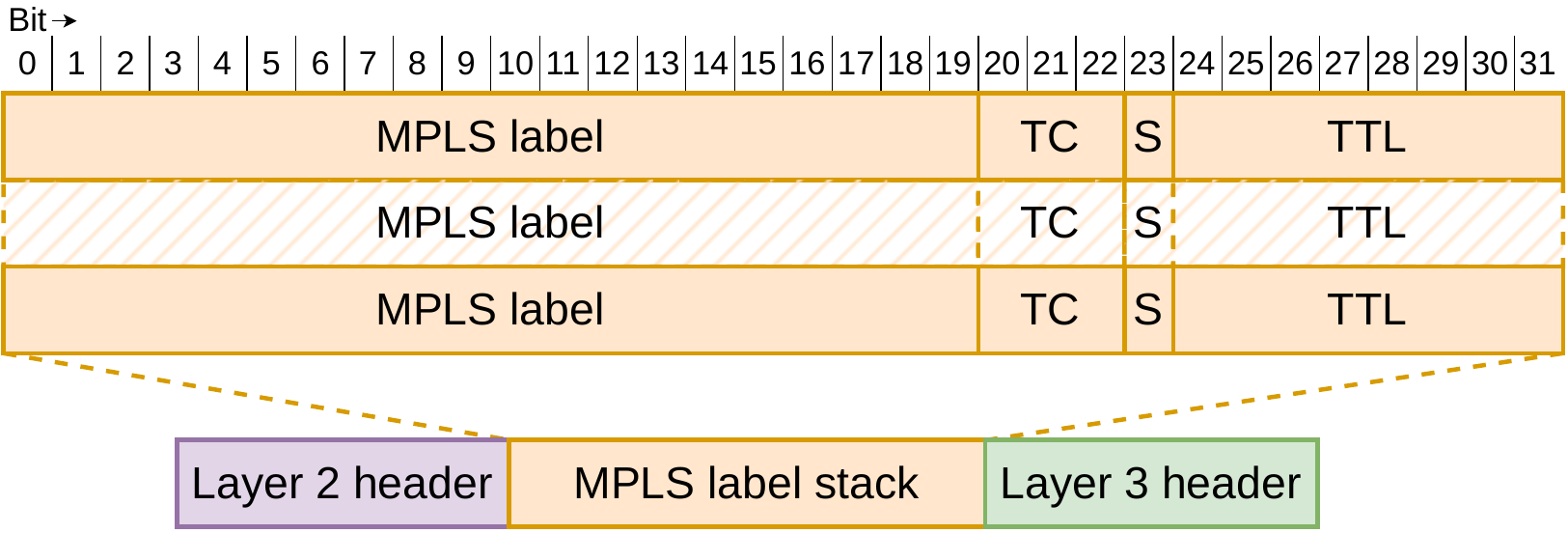}{The \acs{MPLS} label stack is located between the Layer 2 header and the Layer 3 header. An \acs{MPLS} stack consists of one or more \acp{LSE}, each of which consists of a label value, a traffic class (TC), a bottom of stack bit (S), and a time-to-live (TTL) field.}

\subsubsection{MPLS Special Purpose Labels}
MPLS labels do not only contain addressing information but also encapsulate information for special purposes.
A special label value range is reserved for these purposes and must not be used for forwarding~\cite{rfc9017}.
They are called \acfp{bSPL} and indicate an operation to be performed by the \acs{LSR}.
An \acs{MPLS} label stack containing a \acs{bSPL} \acs{LSE} is illustrated in \fig{pdfs/ihle5}.

For \acp{bSPL}, only 16 reserved values are available.
Therefore, the IETF defined the specific \acs{bSPL} label with value $15$, called an extension label, which indicates that the \ac{LSE} following that \ac{bSPL} contains \ac{eSPL} values~\cite{rfc9017}.
For \acp{eSPL}, more label values are reserved and available for extensions.
An \acs{MPLS} label stack containing an \acs{eSPL} \acs{LSE} is illustrated in \fig{pdfs/ihle6}.

\twosubfigepsSideBySide{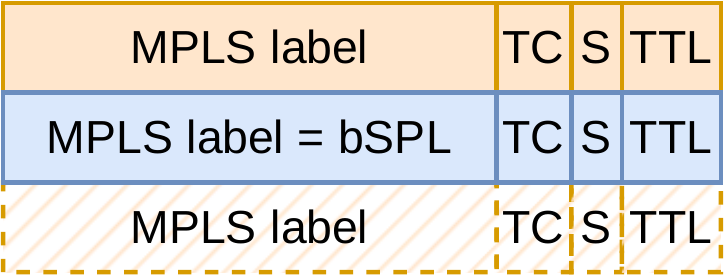}{An \acs{MPLS} label stack with a \acs{bSPL} \acs{LSE}.}{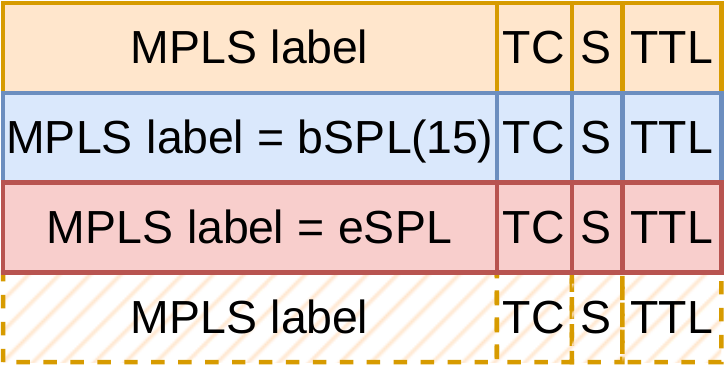}{An \acs{MPLS} label stack with a \acs{bSPL} \acs{LSE} indicating an \acs{eSPL} \acs{LSE}.}{\acs{MPLS} label stacks containing \acs{bSPL} and \acs{eSPL} \acp{LSE}\cite{rfc9017}.}

However, currently only two \acs{eSPL} values are allocated which are used for \acs{SFC} in RFC 8595\cite{rfc8595}.
For these \acs{eSPL} values, it is assumed that no other extension label is present in the MPLS stack\cite{song-mpls-extension-header-13}.
This makes it difficult to combine \acs{SFC} \acp{eSPL} with other extensions such as the \acf{AMM}.
Further, MPLS extensions often require modifications to existing specifications\cite{ietf-mpls-mna-fwk-05}.

\subsection{The MNA Framework}
\label{sec:header}
This section summarizes the concept of the \acs{MNA} framework, including an overview, the header encoding for network actions, and scopes in \acs{MNA} proposed by the IETF MPLS working group. 

\subsubsection{Overview}
\revieweri{The \acs{MNA} framework provides a general mechanism for the transmission and processing of predefined network actions and their required data to facilitate extensions to the MPLS protocol~\cite{ietf-mpls-mna-fwk-05}.
To that end, the working group proposed a new header encoding for network actions in the \acs{MNA} framework\cite{ietf-mpls-mna-hdr-04}.
Network actions are either located in-stack or post-stack.
In this work, we consider in-stack network actions.
Generally, a network action in \acs{MNA} contains an opcode that identifies the operation that the \acs{LSR} will perform on the packet.
Those network actions are encoded as \acp{LSE} in the \acs{MPLS} stack.
A packet can contain multiple network actions.}

\revieweri{A network action may require input parameters to process the indicated network action or may write the result from the processed network action into the packet header.
This data is carried in the \acs{MPLS} stack leveraging the \acs{MNA} encoding and is called \acf{AD}.}
\subsubsection{The MNA Header Encoding}
\label{sec:encoding}
In \acs{MNA}, a stack of related \acp{LSE} in the MPLS stack containing network actions and \acs{AD} is called a \acf{NAS}\cite{ietf-mpls-mna-fwk-05, ietf-mpls-mna-hdr-04}.
A \acs{NAS} is inserted below a forwarding label in the \acs{MPLS} stack. 
Multiple \acs{NAS} can exist in an \acs{MPLS} stack.
\fig{pdfs/ihle7} shows an example of two \acs{NAS} inserted into an \acs{MPLS} stack with stacked forwarding labels.

A \acs{NAS} is comprised of multiple \acp{LSE} with different purposes.
Essentially, a \acs{NAS} is a \acs{bSPL} \acs{LSE} followed by network action \acp{LSE}.
For network action \acp{LSE}, the traditional \acs{MPLS} \acs{LSE} encoding is repurposed.
A \acs{NAS} consists of four differently encoded \acp{LSE} listed below\cite{ietf-mpls-mna-hdr-04}.

\begin{itemize}
    \item The \acs{NAS} indicator \acs{LSE} (Format A),
    \item the initial opcode \acs{LSE} (Format B),
    \item the subsequent opcode \acs{LSE} (Format C),
    \item the \acf{AD} \acs{LSE} (Format D).
\end{itemize}

\figeps[\columnwidth]{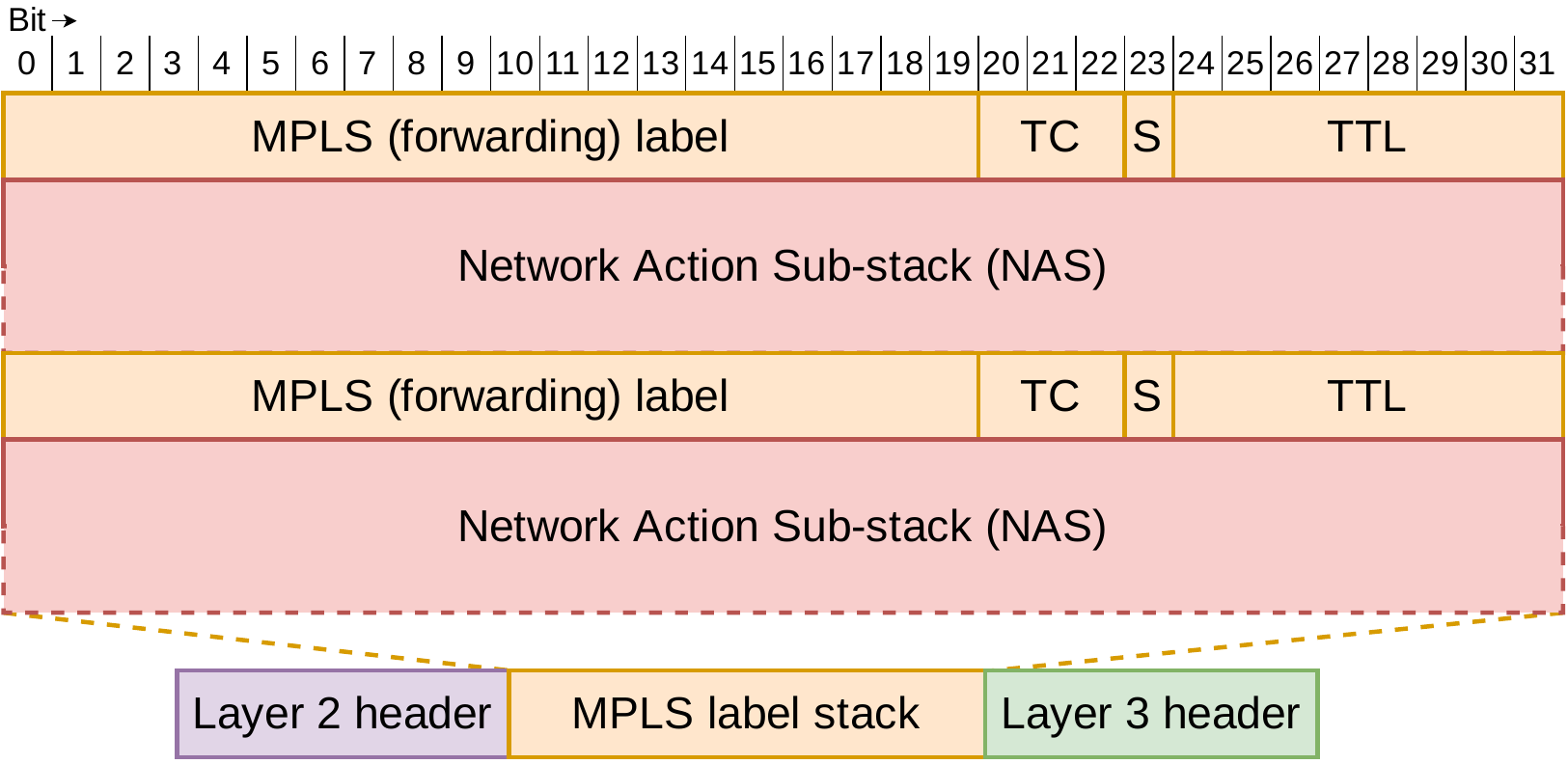}{An example of two \acs{NAS} inserted into an \acs{MPLS} label stack. A \acs{NAS} consists of at least two \acp{LSE} and up to 17 \acp{LSE}.}

For network actions and \acs{AD} \acp{LSE}, the encoding of an \acs{MPLS} \acs{LSE} is repurposed as shown in \fig{pdfs/ihle8}.

\figeps[\columnwidth]{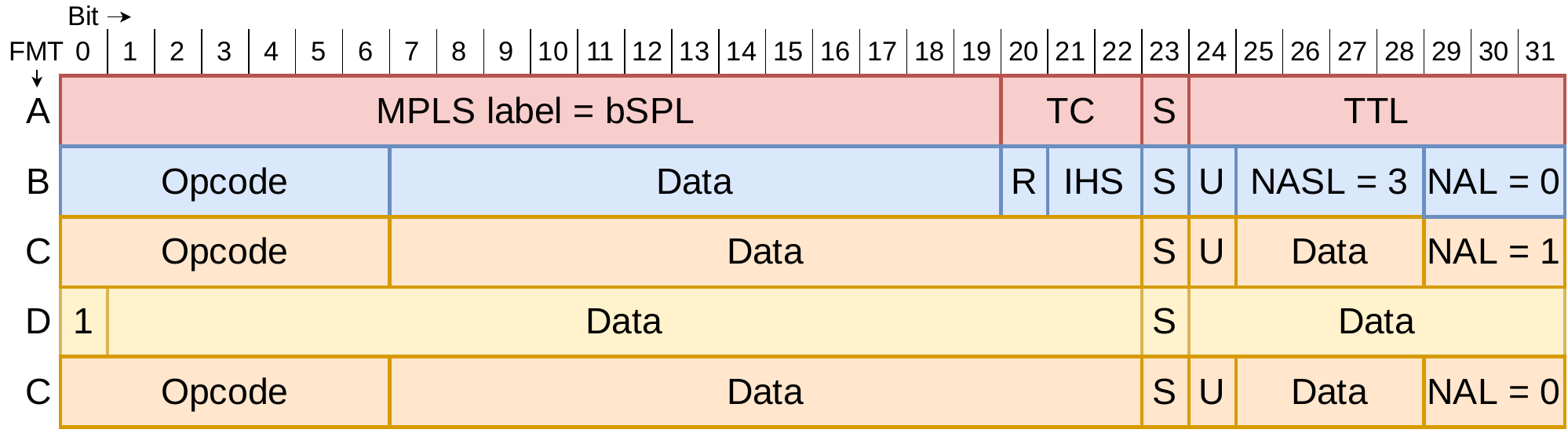}{The four different \acs{LSE} encodings of a \acf{NAS} in the \acs{MNA} framework.}

In the following, the different encodings are explained.
A \acs{NAS} starts with a \acs{bSPL} value (\textit{Format A}) that indicates the beginning of the \acs{NAS}.
This \acs{LSE} is referred to as \acs{NAS} indicator.
Following the \acs{NAS} indicator, the mandatory initial opcode \acs{LSE} (\textit{Format B}) defines the first network action to be processed in this sub-stack.
The repurposed fields in the initial opcode format are shown in \fig{pdfs/ihle8} and are briefly described below\cite{ietf-mpls-mna-hdr-04}.
\begin{itemize}
    \item Opcode: the operation code that identifies this network action.
    \item Data: \acs{AD} belonging to this network action, e.g., flags. 13 bits are available for \acs{AD} in Format B.
    \item IHS: the scope of this sub-stack. This can be either ingress-to-egress (\textbf{I}2E), hop-by-hop (\textbf{H}BH), or \textbf{S}elect. More on this in \sect{placement}.
    \item \acf{NASL}: the number of additional network actions and \acs{AD} \acp{LSE} in this \acs{NAS} excluding the Format B LSE, i.e., the length of this \acs{NAS}.
    \item \acf{NAL}: the number of \acs{AD} \acp{LSE} following and belonging to this network action.
\end{itemize}
The \acs{NASL} and the IHS field defined in the Format B \acs{LSE} are valid for the entire \acs{NAS}.
The R, S, and U fields are not relevant in this work.
The bare minimum \acs{NAS} consists of the \acs{NAS} indicator (Format A) \acs{LSE} and the initial opcode (Format B) \acs{LSE}.
Only one Format B \acs{LSE} is allowed in an \acs{NAS}.

Optionally, subsequent opcode \acp{LSE} (Format C) or \acs{AD} \acp{LSE} (Format D) follow.
The \textit{Format C} \acs{LSE} is a simplified version of the Format B encoding.
The \textit{Format D} \acs{LSE} contains \acs{AD} which relates to the preceding Format B, or Format C \acs{LSE}.
\acs{AD} can be carried either as part of a Format B or Format C \acs{LSE}, or as a separate \acs{AD} \acs{LSE} in Format D.
In the Format C \acs{LSE}, 20 bits are available for opcode-specific \acs{AD}.
Finally, the Format D \acs{LSE} contains $30$ bits for opcode-specific \acs{AD}.
The semantics of the \acs{AD} are left to the predefined network action.
A \acs{NAS} can contain up to 16 network actions and \acs{AD} \acp{LSE}.
Together with the \acs{NAS} indicator, a \acs{NAS} therefore contains up to 17 \acp{LSE}.
After exposing a \acs{NAS} to the top, it must be popped.

\subsubsection{MNA Scopes and MNA with SR-MPLS}
\label{sec:placement}
Network actions in the \acs{MNA} framework can be processed on selected nodes, on the egress node only, or on all nodes along a path.
To that end, a \acs{NAS} specifies the scope of the contained network actions in the IHS field of the Format B \acs{LSE}.
Available scopes in the proposed \acs{MNA} header encoding are \textit{select}, \textit{\acf{I2E}}, and \textit{\acf{HBH}}\cite{ietf-mpls-mna-hdr-04}.
In the following, the three scopes are described.
Then, an example for the placement of differently scoped \acs{NAS} in the MPLS stack is given in \fig{pdfs/ihle9}.

A \textit{select}-scoped \acs{NAS} is processed by one specific node on the path.
It is located below the forwarding label for the specific node.
Only the node that exposes this \acs{NAS} to the top of stack, i.e., that pops the preceding \acs{MPLS} forwarding label, processes the select-scoped NAS.
This \acs{NAS} is popped after processing.

The \textit{\acs{I2E}}-scoped \acs{NAS} is only processed by the egress \acs{LER}.
The \acs{I2E} scope provides ingress to egress transport of network actions and is a separate scope to provide data from the ingress node specific to the egress node.
It is placed at the bottom of the \acs{MPLS} stack.

\textit{\acs{HBH}}-scoped network actions must be processed on each node along the path from source to destination.
An \acs{HBH}-scoped \acs{NAS} is located deeper in the stack, i.e., below another forwarding label than the top-of-stack label.
Therefore, an \acs{LSR} must search deeper in the MPLS stack to find an \acs{HBH}-scoped \acs{NAS}.
\revieweri{This problem is further described in \sect{eval_signaling}.}

\revieweri{\fig{pdfs/ihle9} illustrates the differently scoped \acs{NAS} and their placement in an exemplary \acs{MPLS} label stack in an \acs{SR}-\acs{MPLS} environment of two \acp{LSR} $R_1$ and $R_2$, and one egress \acs{LER} $R_3$. 
Further, \fig{pdfs/ihle9} shows which nodes process the differently scoped \acs{NAS}.}

\figeps[\columnwidth]{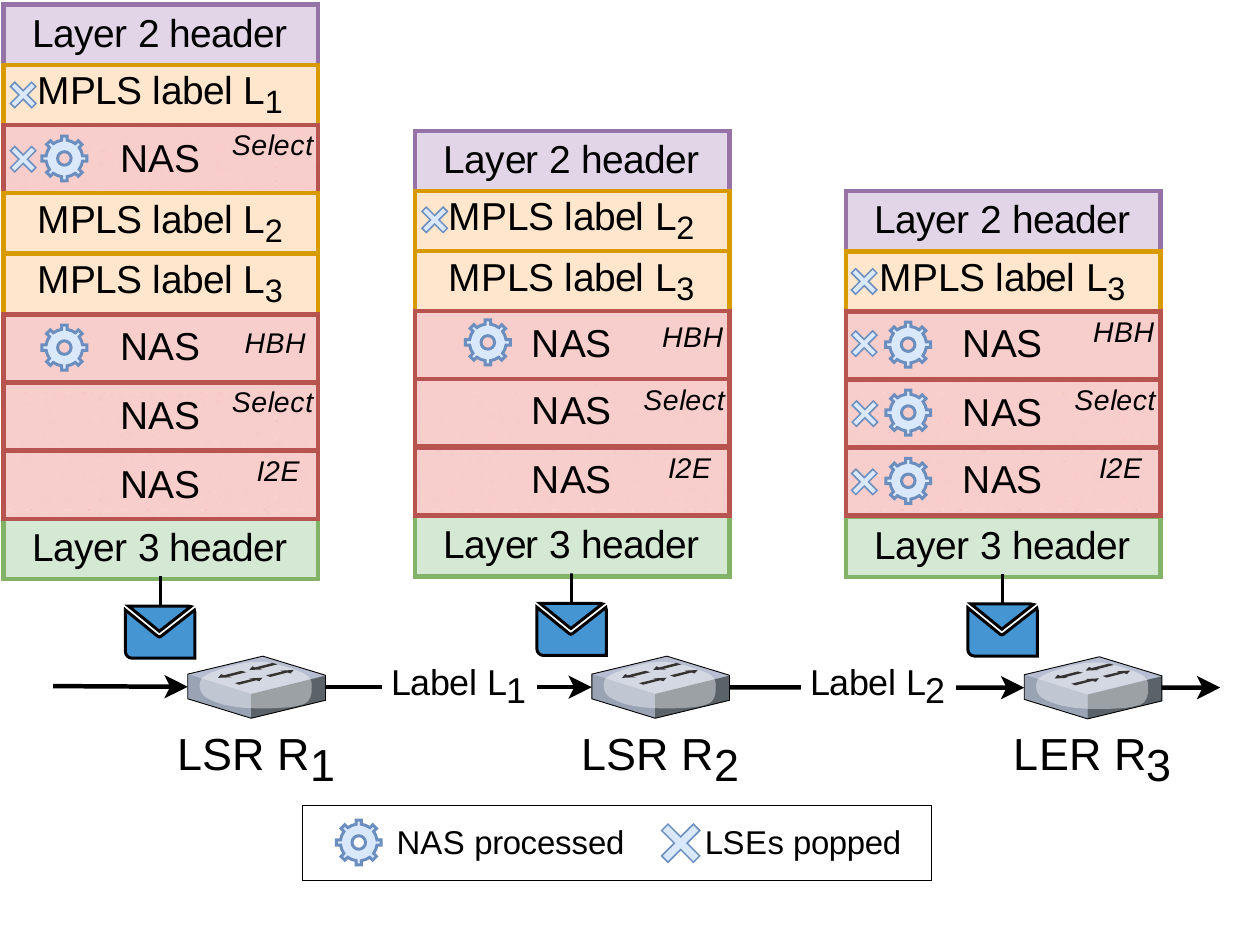}{An example \acs{MPLS} label stack in an SR-MPLS environment with  adjacency segments. The HBH-scoped \acs{NAS} is processed by all nodes, the select-scoped \acs{NAS} are processed by nodes that pop the preceding forwarding label, and the \acs{I2E}-scoped \acs{NAS} is processed by the egress \acs{LER}.}

The MPLS stack in \fig{pdfs/ihle9} encodes MPLS forwarding labels for adjacency segments.
In this example, \acp{LSR} parse the entire MPLS stack.
Therefore, in \fig{pdfs/ihle9}, $R_1$ executes network actions in the select-scoped \acs{NAS} located below the top-of-stack forwarding label, and the \acs{HBH}-scoped \acs{NAS} located below the forwarding label $L_3$.
$R_1$ pops the select-scoped \acs{NAS} after popping $L_1$.
$R_2$ executes network actions in the \acs{HBH}-scoped \acs{NAS}.
$R_3$ executes network actions in the HBH-scoped \acs{NAS}, the select-scoped \acs{NAS}, and the \acs{I2E}-scoped \acs{NAS} located below its forwarding label.
In this example, penultimate hop popping is not applied.
With penultimate hop popping, $R_2$ removes the last forwarding label but must not remove the \acs{NAS} exposed to the top. 
Then, $R_3$ may receive a \acs{NAS} at the top of stack~\cite{ietf-mpls-mna-hdr-04}.
\section{Identified Use Cases For the MNA Framework}
\label{sec:use_cases}
\revieweri{In this section, we describe five use cases identified by the IETF working group\cite{ietf-mpls-mna-usecases-03}, namely no further fast reroute (NFFRR), in-situ \acs{OAM}, \ac{SFC}, \ac{AMM}, and network slicing.
Many of the identified use cases for \acs{MNA} exist as a technology-agnostic mechanism and  are implemented and adapted for the \acs{MPLS} protocol.
\tabl{summary_use_cases} shows a summary of the identified use cases for the \acs{MNA} framework.
For each use case, the technology-agnostic mechanism, the adaption to \acs{MPLS}, and a proposal for the \acs{MNA} framework are shown.
In the following sections, the use cases are described in more detail.}

\begin{table}[htb!]
	\caption{Overview of identified use cases for \acs{MNA}.}
    \begin{tabularx}{\columnwidth}{|X|X|X|X|}
    \hline
    \textbf{Use case}        & \begin{tabular}[c]{@{}l@{}}\textbf{Technology-}\\\textbf{agnostic}\\\textbf{mechanism}\end{tabular} & \begin{tabular}[c]{@{}l@{}}\textbf{Adaption to}\\\textbf{MPLS}\end{tabular} & \begin{tabular}[c]{@{}l@{}}\textbf{Proposal for}\\\textbf{MNA}\end{tabular} \\ \hline \hline
    \textbf{NFFRR}            & \multicolumn{1}{c|}{-}                                                                        & Kompella \textit{et al.}\cite{nffrr}                                 & Saad \textit{et al.}~\cite{li-mpls-mna-nffrr-01}                         \\ \hline
    \textbf{IOAM}            & RFC~9197~\cite{rfc9197}, RFC~9326~\cite{rfc9326}                                & RFC~5586~\cite{rfc5586}, RFC~6669~\cite{rfc6669}                                                                     & Gandhi \textit{et al.}~\cite{mpls-oam}, Mirsky \textit{et al.}~\cite{mb-mpls-ioam-dex-08}                                    \\ \hline
    \textbf{SFC}             & RFC~7665~\cite{RFC7665}, RFC~8300~\cite{RFC8300}                                    & RFC~8595~\cite{rfc8595}                                  & MNA use cases draft~\cite{ietf-mpls-mna-usecases-03}                                                                     \\ \hline
    \textbf{AMM}             & RFC~9341~\cite{rfc9341}, RFC~9342~\cite{rfc9342}                                 & RFC~8372~\cite{rfc8372}, Cheng~\textit{et al.}~\cite{ietf-mpls-inband-pm-encapsulation-13}  & Cheng \textit{et al.}~\cite{cx-mpls-mna-inband-pm-04}                                                                     \\ \hline
    \begin{tabular}[c]{@{}l@{}}\textbf{Network}\\\textbf{slicing}\end{tabular} & RFC~9543~\cite{rfc9543}                                            &  Saad \textit{et al.}~\cite{ietf-teas-ns-ip-mpls-04}                                                                    &  Li \textit{et al.}~\cite{li-mpls-mna-nrp-selector-01}                                                                    \\ \hline
    \end{tabularx}
    \label{tab:summary_use_cases}
\end{table}

\subsection{No Further Fast Reroute (NFFRR)}
The \acs{MPLS} \ac{FRR} mechanism defined in RFC 4090\cite{rfc4090} is a well-established mechanism to counter link and node failures in an \acs{MPLS} network.
In \acs{FRR}, backup tunnels are established to bypass a section of an \ac{LSP} in case of a failure.
Kompella \textit{et al.}\cite{nffrr} identified the problem of looping packets in a network with multiple failures where \acs{FRR} is applied multiple times, eventually leading to congestion and packet loss.
An example is illustrated in \fig{pdfs/ihle10} and explained below.

\figeps[\columnwidth]{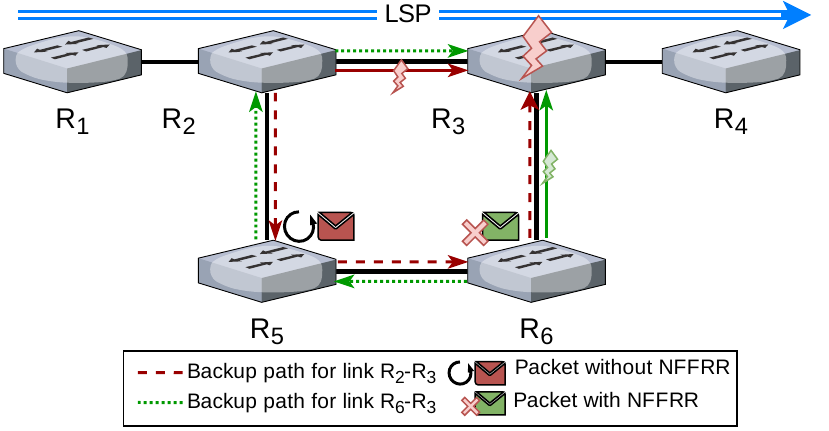}{In the network topology, an LSP $R_1$--$R_2$--$R_3$--$R_4$ is established. The link $R_2$--$R_3$ is protected by the path $R_2$--$R_5$--$R_6$, and the link $R_6$--$R_3$ is protected by the path $R_6$--$R_5$--$R_2$.
If the link $R_2$--$R_3$ and $R_6$--$R_3$ fail at the same time, packets loop between the backup paths.
With NFFRR, packets are marked on the first reroute and dropped on a second reroute\cite{nffrr}.}

In \fig{pdfs/ihle10}, an \acs{LSP} from node $R_1$ to node $R_4$ over $R_2$ and $R_3$ is established.
If the links $R_2$--$R_3$ and $R_6$--$R_3$ fail simultaneously, e.g., by a node failure of node $R_3$, a packet from $R_1$ arriving at $R_2$ is rerouted using the backup tunnel $R_2$--$R_5$--$R_6$.
Once the packet arrives at node $R_6$, the packet is again rerouted using the backup tunnel $R_6$--$R_5$--$R_2$ because the link $R_6$--$R_3$ is down.
The packet is rerouted to node $R_2$ where it is again rerouted to node $R_6$.
Packets loop between $R_2$, $R_5$, and $R_6$ until the TTL expires.
This leads to congestion of the links $R_2$--$R_5$ and $R_5$--$R_6$.

As a solution, Kompella \textit{et al.} propose that an \acs{LSR} adds a \acs{bSPL} \acs{MPLS} label if a packet was rerouted via a bypass \acs{FRR} tunnel and if another bypass via a \acs{FRR} tunnel is not desired.
A packet carrying such a \acs{bSPL} allows an \acs{LSR} to distinguish fast-rerouted packets from regular packets.
The \acs{LSR} that encounters such an \acs{bSPL} \acs{MPLS} label must not send the packet over a \acs{FRR} tunnel.
\FI{With this approach, node $R_2$ marks the packet in \fig{pdfs/ihle10} with the NFFRR \acs{bSPL}. 
Node $R_6$ does not reroute the packet back to node $R_2$ but drops it because the NFFRR \acs{bSPL} label is present.}

However, the number of available \acp{bSPL} is limited.
Therefore, Saad \textit{et al.}\cite{li-mpls-mna-nffrr-01} propose a network action to leverage the \acs{MNA} framework for NFFRR.
They propose to indicate whether a packet was rerouted with a single bit in the \acs{AD} of a network action.

\subsection{In-Situ OAM}
\label{sec:oam}
In-situ OAM (IOAM) collects operational and telemetry information as packets traverse a specific path.
This information can be leveraged for traffic engineering and monitoring.
IOAM does not send probe packets to collect metrics. 
Instead, the information is added to or the collection is triggered by existing data packets.

\FI{IOAM data can be collected in two ways.
First, by adding information to each packet and evaluating the information at the tail-end.
Second, by exporting information from packets received by the \acs{LSR} to a collector in the network, e.g., the control plane.
The first method is called the passport mode and the second is called the postcard mode.
The methods are illustrated in \fig{pdfs/ihle11_full}~\cite{rfc9326}.}

\twosubfigeps{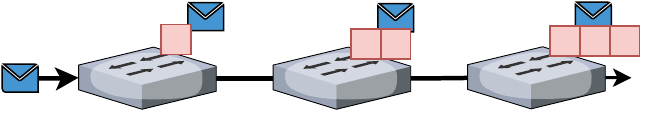}{Passport mode.}{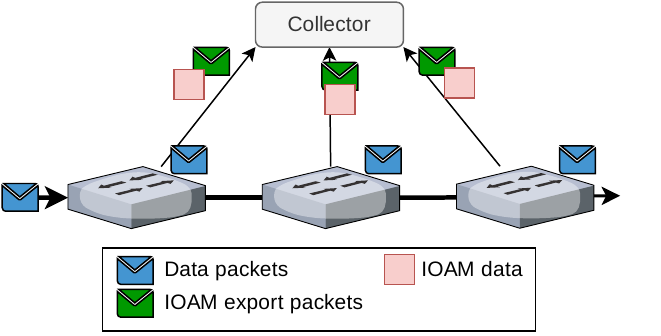}{Postcard mode.}{Operation modes in IOAM.}

In the passport mode in \fig{pdfs/ihle11}, each \acs{LSR} places a \textquote{stamp}, i.e., the IOAM data, on the \textquote{passport}, i.e., the header, of a packet.
The passport mode is used in RFC 9197~\cite{rfc9197}, e.g., for path tracing.
RFC 9326~\cite{rfc9326} introduces the direct export option which leverages the postcard mode for data collection.
In the postcard mode in \fig{pdfs/ihle12}, each \acs{LSR} \textquote{writes a postcard}, i.e., sends a message containing the IOAM data to the collector.
The information is either exported as raw data per packet or aggregated, processed, and exported using protocols such as IPFIX.
No information is added to the data packet with the postcard mode.
Which mode to use and what data to collect is indicated in the packet, e.g., with bSPL labels in MPLS\cite{rfc5586}.




Gandhi \textit{et al.}~\cite{mpls-oam} propose an approach to encapsulate IOAM data in \acs{MPLS} data planes.
They leverage the \acs{MNA} framework for this approach.
Their draft encapsulates the IOAM data from RFC 9197 and RFC 9326 in a post-stack network action.
\FI{Mirsky \textit{et al.}~\cite{mb-mpls-ioam-dex-08} propose an alternative approach that encapsulates the direct export option from RFC 9326 in an in-stack network action.
}


\subsection{Service Function Chaining (SFC)}
\ac{SFC} steers packets through a defined set of network functions, such as firewalls, IDS, and NAT.
Packets are classified at the network edge and the classification decision is stored in an \acs{SFC} header.
This header is leveraged to direct packets to the desired network functions for processing\cite{RFC7665}.
\FI{To that end, SR-MPLS or the \acf{NSH} can be used\cite{HaSt22}.
An example of a \acs{SFC} using SR-MPLS is given in \fig{pdfs/ihle13}.}

\figeps[\columnwidth]{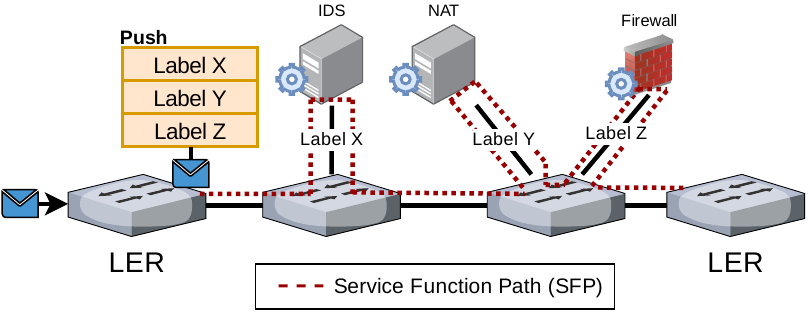}{A packet is classified at the ingress LER. Based on the classification, an SR-MPLS stack is pushed onto the packet that encodes the service function path (SFP). This path directs the traffic to various service functions.}

In \fig{pdfs/ihle13}, the ingress \acs{LER} classifies a packet and adds an SR-MPLS stack which encodes the service function path (SFP).
The SFP steers the traffic to different service functions.
Alternatively to SR-MPLS, the \acs{NSH} can be used to direct traffic to the service function.
The \ac{NSH} defined in RFC 8300\cite{RFC8300} describes an encapsulation for \acp{SFC}.

RFC 8595\cite{rfc8595} introduces an \acs{MPLS}-based forwarding plane for \acs{SFC} leveraging \acs{eSPL} values.
Two consecutively stacked \acs{MPLS} labels encode the \acs{NSH} in an \acs{MPLS} label stack.
The fields of an \acs{LSE} are repurposed to reflect the data encoded in an \acs{NSH}.
However, this imposes some limitations on the \acs{NSH}. 
Because the 24-bit wide service path identifier of an \acs{NSH} is mapped to the 20-bit wide \acs{MPLS} label field, the MPLS representation of \acs{NSH} must not assign values that exceed the size of an \acs{MPLS} label.
Furthermore, the eight-bit wide service index, which indicates the location of the service function within the path, is mapped to the second \acs{MPLS} label, leaving 12 bits unused.


The proposed \acs{MNA} framework is a generalization of the concept described in RFC 8595. 
In \acs{MNA}, the \acs{NSH} can be encoded more efficiently in the \acs{MPLS} label stack without imposing restrictions.
To that end, the \acs{eSPL} values can be introduced as network action opcodes.
The service-path identifier, the service index, and metadata can be encoded as \acs{AD} in a \acs{NAS}.

\subsection{Performance Measurement with Alternate-Marking Method (AMM)}
\label{sec:amm}
The \acf{AMM} defined in RFC 9341~\cite{rfc9341} and RFC 9342~\cite{rfc9342} is a mechanism to measure packet loss and delay.
In general, packet loss can be detected using sequence numbers in packets.
However, this requires the insertion of sequence numbers into packets.
Further, devices must be able to extract and verify the sequence number~\cite{rfc9341}.
\acs{AMM} is an alternative approach to measure the packet loss and delay between two points.
For packet loss measurement, the number of sent packets is counted on the sending end and the number of received packets is verified on the receiving end.
If the number of sent packets is not equal to the number of received packets, packets are lost.
\acs{AMM} requires synchronization of both sides, i.e., they must be referring to the same set of packets.
For this purpose, a flow is divided into batches by marking packets of the same batch and the same flow with the same color.
For each consecutive batch, the color is alternated.
On changing the color, the number of sent packets in the previous batch is exported via an out-of-band channel, e.g., to the control plane.
This allows the control plane to verify the number of packets of the same color.
The number of packets in a batch is either fixed or based on the packet rate and a fixed timer.
For delay measurement, the timestamp of the color change event is exported to the control plane.
\FI{An example using \acs{AMM} for packet loss measurement is given in \fig{pdfs/ihle14}.}

\figeps[\columnwidth]{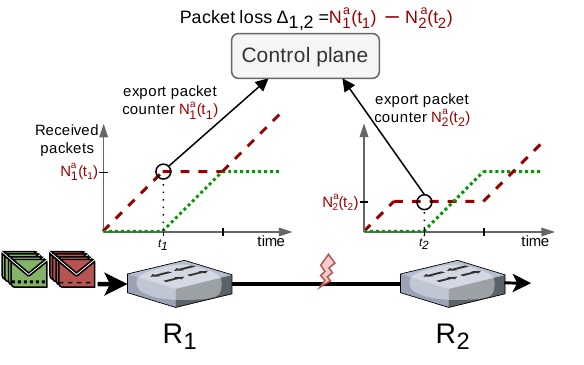}{Packets are sent in colored batches. Each \acs{LSR} counts the packets per color. If the color alternates, the packet counter is exported to the control plane. The control plane calculates packet loss based on the received counters.}

\revieweri{In \fig{pdfs/ihle14}, colored packet batches are sent over an \acs{LSP} of two \acp{LSR} $R_1$ and $R_2$.
The color-specific packet counters per \acs{LSR} over time are shown in the plots.
At $t_{1}$ the color alternates in \acs{LSR} $R_1$, and at $t_{2}$ in \acs{LSR} $R_2$.
Therefore, the \acp{LSR} $R_1$ and $R_2$ export their counter values of color $a$, $N^a_1(t_1)$ and $N^a_2(t_2)$, to the control plane.
The control plane then calculates the packet loss on the link from node $R_1$ to $R_2$ of this batch $\Delta_{1,2} = N^a_1(t_1) - N^a_2(t_2)$.}

Packets are colored by setting a bit in the header, such as the drop eligible indicator bit for Ethernet frames. 
Other header fields may be used depending on the application.
Cheng \textit{et al.}~\cite{ietf-mpls-inband-pm-encapsulation-13} describe an approach to leverage \acs{eSPL} entries to implement \acs{AMM} in \acs{MPLS} networks.
In~\cite{cx-mpls-mna-inband-pm-04}, Cheng \textit{et al.} describe another approach to leverage the \acs{MNA} framework to implement \acs{AMM} in \acs{MPLS} networks.
Here, a \acs{NAS} with the \acs{AMM} network action is described.
\revieweri{The encoding proposed by \cite{cx-mpls-mna-inband-pm-04} for an \acs{AMM} network action in \acs{MNA} is shown in \fig{pdfs/ihle15}.}

\figeps[\columnwidth]{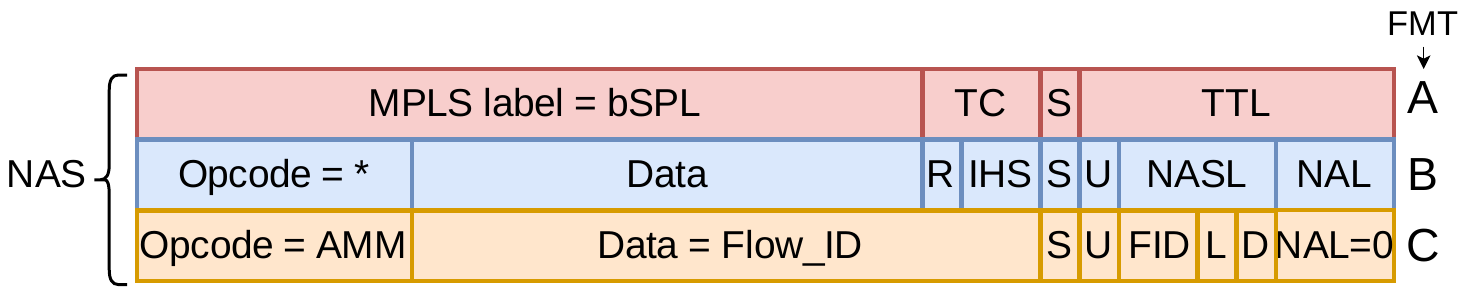}{The AMM network action is indicated by an opcode in the Format C \acs{LSE}. The data field of the network action contains the flow ID and the packet color\cite{cx-mpls-mna-inband-pm-04}.}

\acs{AMM} operates on a per-flow basis.
The 18-bit wide flow ID is embedded into the data field of the network action shown in \fig{pdfs/ihle15}.
The network action including its \acs{AD} does not fit into a Format B \acs{LSE} and requires a Format C \acs{LSE}.
The flow ID is distributed across the 16 bits of the first data field and the two most significant bits of the second data field.
The two least significant bits of the data field indicate the color for packet loss measurement (L) and delay measurement (D).
This network action is implemented in P4-MNA for packet loss measurement and evaluated in \sect{amm_eval}.
\subsection{Network Slicing}
\label{sec:network_slicing}
An IETF network slice, defined in RFC 9543~\cite{rfc9543}, provides connectivity coupled with a set of specific commitments of network resources, such as reserved bandwidth or latency.
Network slices provide end-to-end logical networks over a shared physical infrastructure for various applications, such as 5G networks and VPNs.
They act as an overlay network, creating multiple isolated virtual networks.
A network slice may stretch across multiple domains of a provider.
The configuration of a network slice, such as nodes, links, buffers, queuing resources, and scheduling resources, is described in a \ac{NRP}.
An \acs{NRP} must be configured prior to operation, e.g., with netconf, or via IGP signaling\cite{ietf-teas-ns-ip-mpls-04}.
Each \acs{NRP} is identified by an \acs{NRP} selector which is added to each packet.
The \acs{NRP} selector indicates which logical network, i.e., network slice, a packet belongs to.
An \acs{NRP} selector consists of one or more fields in the packet, e.g., an IPv6 address, or an MPLS label\cite{ietf-teas-ns-ip-mpls-04}.
A switch must perform operations on a packet based on the \acs{NRP} selector, e.g., traffic shaping to enforce bandwidth reservations.
An example network topology with two \acp{NRP} sharing a physical infrastructure is shown in \fig{pdfs/ihle16}.

\figeps[\columnwidth]{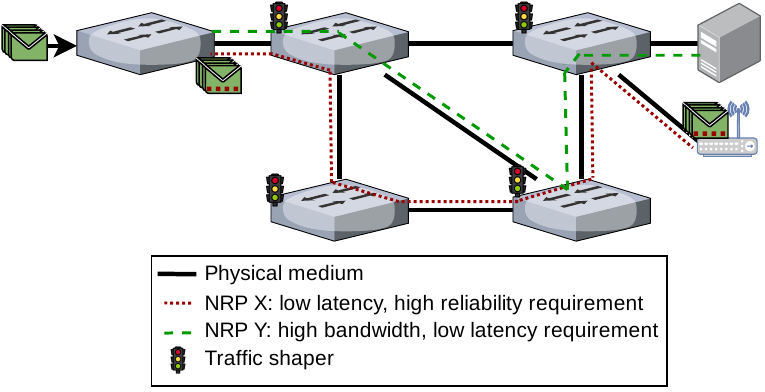}{Two \acp{NRP} sharing the same physical infrastructure. They have different traffic requirements that are enforced by the \acp{LSR}.}

Two \acp{NRP} are configured in the network in \fig{pdfs/ihle16}.
NRP X (dotted) has low latency and high-reliability requirements, e.g., for 5G networking.
NRP Y (dashed) has high bandwidth and low latency requirements, e.g., for video streaming.
The requirements, e.g., bandwidth reservations, are configured in the \acp{LSR}.
Both NRPs share the same physical medium but their traffic is isolated in two logical networks.
At the ingress node, traffic is classified to identify to which \acs{NRP} a packet belongs and a \acs{NRP} selector is pushed to the packet.
Each transit node in the network enforces the \acs{NRP} by traffic shaping and scheduling.

In \cite{ietf-teas-ns-ip-mpls-04}, Saad \textit{et al.} describe an approach for implementing network slices in \acs{MPLS} networks.
Here, \acs{LER} classify incoming traffic and push an \acs{NRP} selector label onto the \acs{MPLS} label stack.
With the \acs{MNA} framework, the \acs{NRP} selector can be carried as a network action in the \acs{MPLS} stack.
\revieweri{Li \textit{et al.} propose an encoding for the NRP selector in a network action which is shown in \fig{pdfs/ihle17}~\cite{li-mpls-mna-nrp-selector-01}.}

\figeps[\columnwidth]{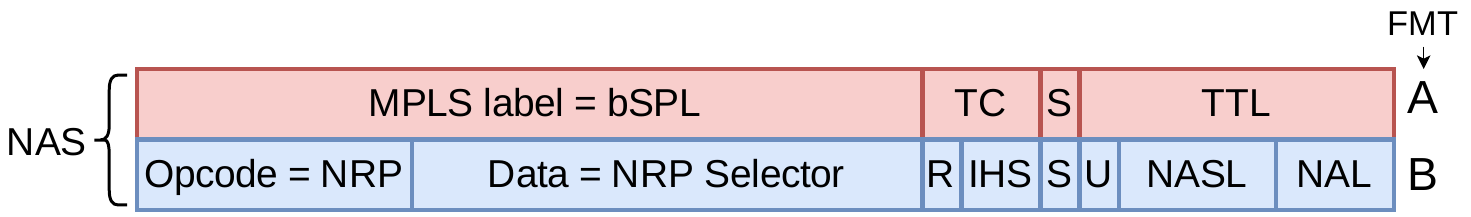}{The NRP network action is indicated by an opcode in the Format B \acs{LSE}. The data field of the network action contains the NRP selector~\cite{li-mpls-mna-nrp-selector-01}.}

\revieweri{A network action for bandwidth reservations with network slicing is implemented in P4-MNA and evaluated in \sect{eval_nrp}.}

\section{Related Work}
\label{sec:related_work}
First, we compare the concept of IPv6 \acp{EH} with the MNA framework.
Then, we review related work that identifies shortcomings in IPv6 \acp{EH}.
Finally, we review implementations of IPv6 \acp{EH}.

\subsection{Comparison of IPv6 Extension Headers with MNA}
\label{sec:ipv6eh}
RFC 2460\cite{rfc2460} defines IPv6 \acp{EH} to expand the functionality of IPv6.
IPv6 \acp{EH} contain options that indicate special operations such as fragmentation or allow for \acf{SR} over IPv6 (SRv6).
\revieweri{
\tabl{comparison_ipv6} compares the concepts from IPv6 \acs{EH} with the MNA framework.}

\begin{table}[htb!]
	\caption{Comparison of IPv6 EH and MNA.}
    \begin{tabularx}{\columnwidth}{|X|X|X|}
\hline
                              & \textbf{IPv6 EH}                                                 & \textbf{MNA}                                         \\ \hline \hline
\textbf{Indication}                    & Next header field                                       & Indicator LSE (bSPL) followed by opcodes                       \\ \hline
\textbf{Scopes}                        & hop-by-hop (HBH), destination (I2E)                                                & Select, HBH, I2E                            \\ \hline
\textbf{Upper bound on \#~extensions}  & Unbounded, chaining of extensions via next header field & 16 network actions per NAS, 1 NAS per scope \\ \hline
\textbf{Upper bound on extension size} & $(2^8-1)\ B = 255\ B$ per option from 8-bit length field in TLV          & $4\ B + (2^4-1) \cdot 4\ B = 64\ B$ in a NAS from 4-bit NASL length field            \\ \hline
\reviewerii{\textbf{Deployment}} & \reviewerii{Global}         & \reviewerii{Limited to a domain}            \\ \hline

    \end{tabularx}
    \label{tab:comparison_ipv6}
\end{table}

\revieweri{IPv6 \acp{EH} are added after the IPv6 base header with their presence and purpose indicated by the next header field in the base header.
Similarly, in MNA, network actions are placed after a forwarding label.
However, unlike IPv6, MPLS does not use next header fields.
Instead, network actions are indicated by a label containing a \acs{bSPL} value followed by \acp{LSE} containing opcodes.}

\revieweri{Both mechanisms allow to specify the scope of actions, i.e., on which node an action is executed.
IPv6 EHs can either be processed only by the destination node (using the destination option) or by every node on the path (using the hop-by-hop option).
MNA further enhances this by allowing network actions to be processed on selected nodes in the path, offering more granular control.}

\revieweri{In terms of structure and size, IPv6 EHs provide high flexibility.
They can be concatenated indefinitely by updating the next header field allowing an unbounded number of extensions per packet. 
However, MNA imposes stricter limits.
The MNA framework supports up to 16 network actions per \acs{NAS} with one NAS allowed per scope. 
This bounded approach simplifies implementation but reduces flexibility compared to IPv6 EHs.}

\revieweri{IPv6 EHs support options encoded in a type-length-value (TLV) format with a theoretical maximum of \SI{255}{\byte} per option due to the 8-bit integer width of the length field.
The length field indicates the number of bytes.
In contrast, the MNA framework limits the size to \SI{64}{\byte} per NAS by the 4-bit wide NASL field offering a smaller size bound.
The NASL field indicates the number of \acp{LSE}, i.e., \SI{4}{\byte}, excluding the Format B \acs{LSE}.}

\reviewerii{IPv6 allows for global, inter-domain communication between devices.
This makes it difficult to deploy extensions, such as new options in a \acs{EH}, because devices around the world must implement new extensions to take advantage of them.
In contrast, MPLS is typically deployed by providers as a Layer 2 and Layer 3 technique within their own domain.
This facilitates the deployment of extensions, such as the \acs{MNA} framework or new network actions in the \acs{MNA} framework.
The usefulness of MNA in a provider's domain does not depend on the deployment of MNA in other domains as providers can choose which network actions to support in their own domain.}

\subsection{Analysis of IPv6 Extension Headers}
\label{sec:ipv6_analysis}
\revieweri{While IPv6 EHs appear more flexible due to their unbounded concatenation and larger extension sizes, this flexibility introduces challenges for hardware implementations.
In the following, we discuss related work on those hardware implications for IPv6 \acp{EH}.}

Custura \textit{et al.}\cite{CuSe24} analyze the transit of packets containing IPv6 \acp{EH} on the Internet.
They perform several experiments in which traffic containing different IPv6 \acs{EH} types and sizes is sent to destinations around the world using RIPE Atlas probes.
\revieweri{In their results, the measured traversal rate decreases with an increasing EH size.
The authors of~\cite{LeIu22} identified that the traversal rate is halved with an EH size of \SI{64}{\byte} and significantly lower with an even larger size.}
They conclude that the successful reception of such a packet depends on the type of \acs{EH} it contains, its size, and the transport protocol used.

Gont \textit{et al.}\cite{rfc9098} describe a reason for the limited forwarding of packets containing IPv6 \acp{EH}.
To efficiently process packets at high data rates, the packets must be processed in the fast path of packet forwarding engines, such as in the data plane of hardware implementations.
The large and dynamic header size of IPv6 \acp{EH} requires the lookup engine to inspect deeply into the packet.
Engines that cannot inspect deep enough into a packet to extract all relevant information typically drop the packet.
Another approach used by some packet forwarding engines with limited lookup capabilities, as described in \cite{rfc9098}, is recirculation.
In this approach, one IPv6 \acs{EH} is processed at a time. 
The packet is then looped back to the ingress and processed again until all \acp{EH} have been processed.
By using recirculation, each packet traverses the processing pipeline multiple times.
If not enough resources for recirculation are available, the performance degrades and packets are dropped.

Routers that are unable to inspect deep enough into a packet, or that do not use recirculation,  can process packets in the slow path, i.e., in the control plane software\cite{rfc9098, CuSe24}.
Processing large packet headers in the control plane drastically reduces the performance.
In addition, control plane processing consumes the resources needed to manage the router.
This facilitates denial-of-service attacks such as described in~\cite{NaGa22}.
Here, a \SI{285}{\byte} arbitrary payload is added with multiple IPv6 options that forces the control plane to validate this payload wasting CPU resources and resulting in a denial of service.
Furthermore, the large size of \acp{EH} prevents firewalls from inspecting transport layer information, making stateful filtering intractable~\cite{GaNa22}.
Since there is no easy solution yet, they suggest dropping packets containing certain IPv6 \acp{EH} as a temporary solution.
They further emphasize that this problem does not arise from a vendor or manufacturer issue but rather from a flaw in the protocol design that allows such large headers to be created.



With IPv6 \acp{EH} being standardized and deployed for many years, it has been proven~\cite{GaNa22, NaGa22, HeVe17,rfc9098,CuSe24} that implementations cannot efficiently support \acp{EH} to their full extent.
The protocol design problem of IPv6 \acs{EH} lies in the large header stacks resulting from chained \acp{EH} and bloated options~\cite{HeVe17}.
They cannot be processed efficiently on hardware.
IPv6 \acp{EH} can be chained arbitrarily with large payloads.
In the \acs{MNA} framework, network actions are stacked.
However, with \acs{MNA}, the maximum size of network actions and their data is limited by the protocol design.
\revieweri{
While the MNA framework is more constrained regarding the structure, number, and size of network actions, it is also more feasible for hardware implementations.
This facilitates the implementation of MPLS extensions in the future.
However, processing many network actions in the fast path remains a challenge due to limited hardware resources.
}
In this work, we therefore investigate how many network actions can be processed in the fast path on hardware, i.e., in the data plane.

\subsection{Implementations of IPv6 Extension Headers}
\label{sec:rel_impl}
\revieweri{In this section, we provide an overview of existing implementations of IPv6 \acp{EH} in both hardware and software.
To the best of our knowledge, no other implementation of the MNA framework than P4-MNA currently exists.}

\revieweri{The authors of \cite{LiLv22} provide a programmable segment routing over IPv6 (SRv6) processor for SFC which is based on a FPGA development board.
The implementation includes SRv6 processing and SFC encapsulation.
Their evaluation shows that their implementation achieves an SRv6 throughput of \SI{100}{\gbps} for \SI{1500}{\byte} packets.
However, the implementation achieved approximately \SI{70}{\gbps} for \SI{128}{\byte} packets.
In \cite{LeBo17}, the authors provide an implementation of SRv6 in the Linux kernel, i.e., a software-based implementation.
Their implementation supports SRv6 and an HMAC TLV option.
They tested their implementation in a \SI{10}{\gbps} testbed.
The authors of \cite{IuDo20} provide an IOAM implementation encapsulated in IPv6 for the Linux kernel.
Their software-based approach achieved approximately \SI{4}{\gbps} for \SI{78}{\byte} packets and \SI{41}{\gbps} for \SI{1236}{\byte} packets.
A whitepaper in \cite{Ta19} evaluates a software-based implementation of SRv6 using the VPP framework.
They achieve a forwarding rate of \SI{48.43}{\gbps} using \SI{192}{\byte} packets.
Additionally, they present an FPGA-based implementation that achieves up to \SI{96.84}{\gbps}.
Software-based implementations, while flexible, are typically limited by general-purpose CPUs and cannot achieve the throughput required for high-speed transit networks operating at hundreds of gigabits per second.
}

\revieweri{While existing implementations of IPv6 \acp{EH} achieve up to \SI{100}{\gbps} under certain conditions, the P4-MNA implementation presented in this paper achieves line rate processing of \SI{400}{\gbps} per port for both small (\SI{128}{\byte}) and large (\SI{1500}{\byte}) packet sizes.
The P4-MNA implementation is capable of processing 32 network actions which are closely related to options in IPv6 \acp{EH} such as IOAM and SFC.}
\section{Introduction to the P4 Programming Language}
\label{sec:p4}
In this section, we give an introduction to the programming language P4.

\ac{P4} is a domain-specific high-level programming language for describing the data plane of programmable switches. 
P4 can be used to implement user-defined algorithms for packet manipulation and forwarding decisions.
A P4 program can be compiled for different targets that implement a specific architecture. 
Such architectures can be either software-based, like the simple\_switch in the BMv2\cite{bmv2}, or hardware-based, like the Intel Tofino™ 2 switching ASIC.

A P4 program consists of a programmable packet parser, several programmable control blocks, and a programmable packet deparser that are arranged sequentially in a pipeline.
The pipeline of the Intel Tofino™ 2 switching ASIC is illustrated in \fig{pdfs/ihle18} and the components are further described in the following.

\figeps[\columnwidth]{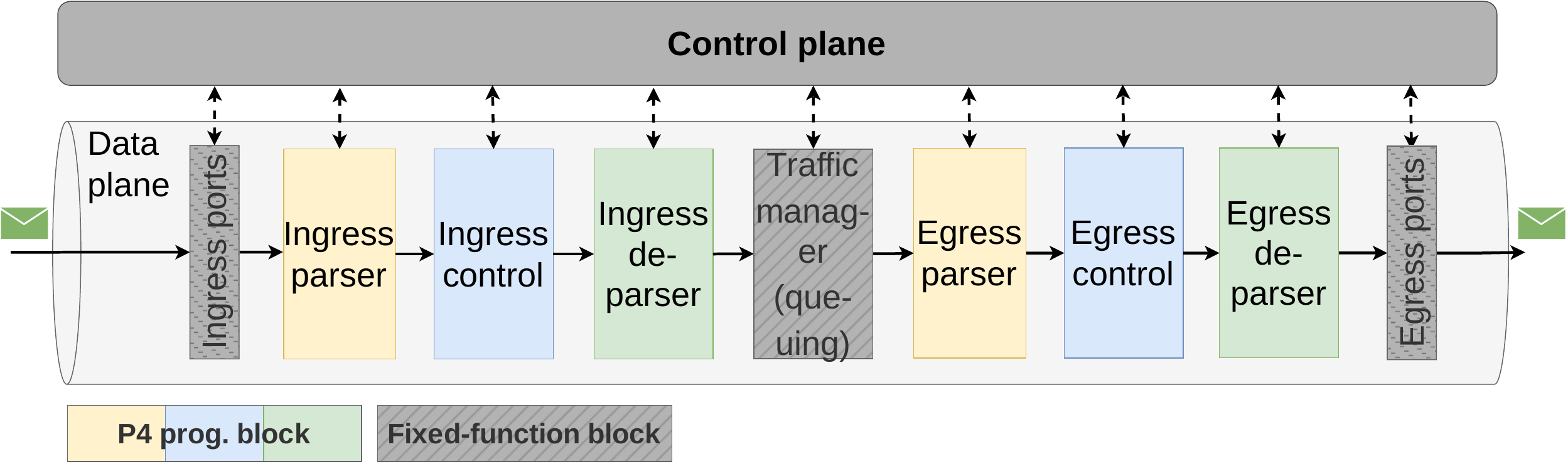}{The pipeline of the Intel Tofino™ 2 switching ASIC consists of a programmable packet parser, control blocks, and a programmable packet deparser for ingress and egress control\cite{tna}.}

A P4 program specifies user-defined metadata and packet headers.
User-defined metadata store values during packet processing and do not exit the switch.
It is comparable to variables in other languages.
User-defined packet headers describe the packet headers that are accessible in the P4 program.
They are parsed by the P4 parser and can be manipulated during pipeline processing.

The P4 parser is modeled as a finite state machine and extracts header information from the packet according to the user-defined parser states.
To that end, user-defined packet headers are defined in the P4 program reflecting the header fields, e.g., an \acs{MPLS} \acs{LSE}.
Multiple of these headers can be aggregated to form a header stack of that specific header type, e.g., an \acs{MPLS} label stack, or a \ac{NAS}.
Entries in a header stack must all share the same header field structure, i.e., different encodings are not possible in a single header stack.
A header stack in P4 is comparable to an array in other languages.
As is usual with arrays, the maximum memory required for an array must be allocated before runtime and cannot be changed dynamically.
Several header stacks and headers can be concatenated to form the complete packet header.
The remaining payload in the packet is ignored and passed on by the P4 program~\cite{p4spec,tna}.

During parsing, bytes are parsed into different user-defined headers, e.g., \acs{LSE} encodings.
An \acs{LSE} parsed into the Format B encoding contains more information about the semantics of the \acs{LSE} than an \acs{LSE} parsed as raw bytes.
The knowledge about the semantics gained during parsing facilitates later processing in the control block.

Control blocks in a P4 program contain the logic of the algorithm.
They consist of \acp{MAT} and can make use of simple arithmetic and logical expressions as well as branching constructs to define the packet processing operations.
The principle of \acp{MAT} is illustrated in \fig{pdfs/ihle19}.

\figeps[\columnwidth]{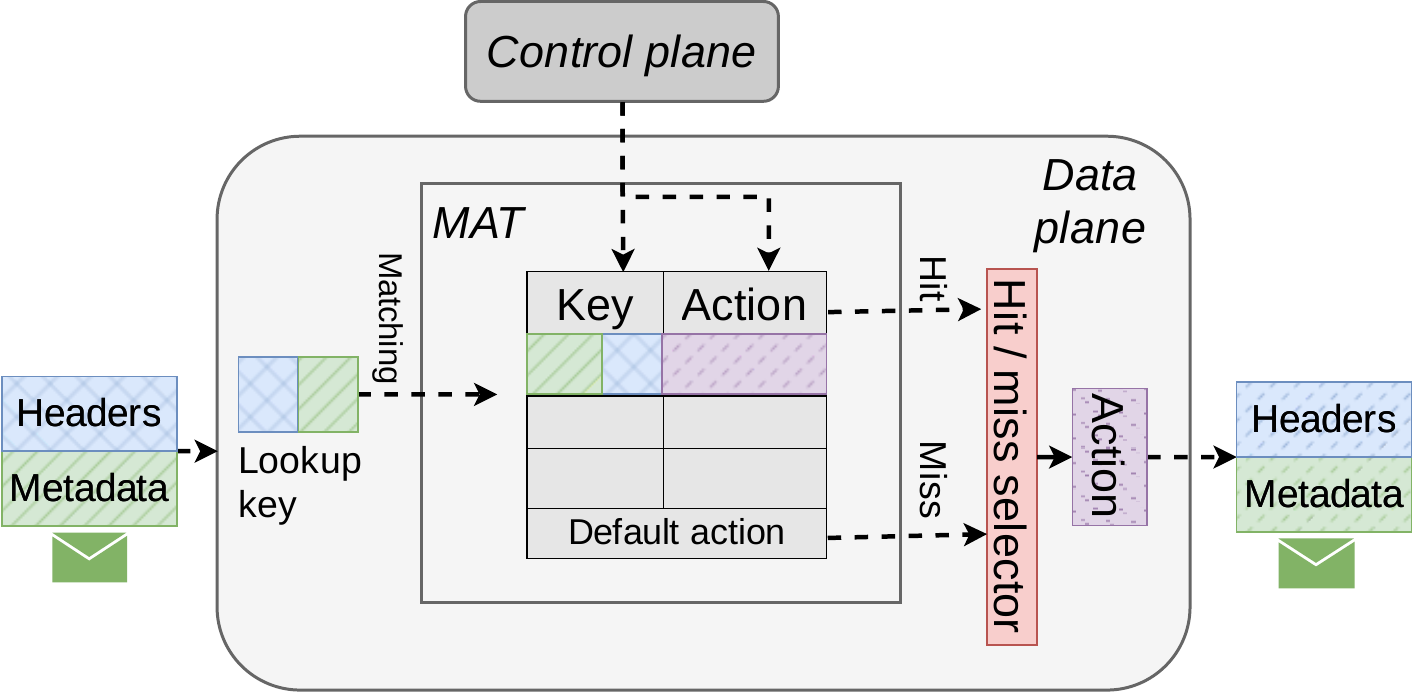}{Selected header fields of a packet form a composite key and are matched in a \acs{MAT}. An associated action is executed. The content of the \acp{MAT} is filled by the control plane\cite{kn}.}

A \acs{MAT} consists of a key definition and an action list.
The key definition comprises selected header fields that are matched in the table.
Upon matching a packet to an entry in the \acs{MAT}, the associated action from the action list is executed.
An action can manipulate packet fields or make a forwarding decision.
While the structure of a \acs{MAT} and the actions are defined in the data plane, the content of the \acp{MAT} is populated by the control plane.

To facilitate line rate processing on hardware, the operations that can be applied during packet processing are limited.
Resubmit is a mechanism to reuse the limited resources of a pipeline.
A resubmit can be triggered in the ingress control block during pipeline processing.
When a resubmit is triggered, a copy of the original packet is sent through ingress processing again.
A resubmit requires no additional bandwidth and adds no processing delay as the packet is not enqueued twice.
A disadvantage of resubmitting is that the packet copy does not contain any changes to the packet headers that were applied during the first pipeline processing.
On the Intel Tofino™, up to eight bytes with metadata can be prepended to a resubmitted packet.
A packet can only be resubmitted once\cite{tna}.
The recirculation mechanism is another mechanism that allows to reuse pipeline resources.
In a recirculation, packets are looped back to the ingress port after processing.
However, recirculation comes at the cost of bandwidth and processing delay which is why we do not consider it in this work.

Externs extend the functionality of P4 with target-specific functions.
Examples of externs are counters, registers, meters, and digest messages.
Registers enable stateful processing of packets.
Values stored in a register are persistent, i.e., they are kept after the packet exits the pipeline.
\revieweri{Meters enable traffic shaping based on a configured rate.}
Digest messages are Intel Tofino™-specific externs that allow small user-defined packets to be sent to the control plane.

More information on P4 can be found in a survey by Hauser \textit{et al.}~\cite{kn}.




\section{P4 Implementation of the \acs{MNA} Framework}
\label{sec:implementation}
This section describes the P4-MNA implementation of the \acs{MNA} framework on the hardware-based Intel Tofino™ 2 switching ASIC using \acf{ISD}.
First, the general architecture of the P4-MNA implementation is described.
Then, we describe how the complex header encoding of the \acs{MNA} framework is parsed in P4.
Next, we explain the processing of network actions in the P4-MNA pipeline.
Finally, we describe the implementation of example network actions for performance measurement using \acs{AMM} and for network slicing.

\subsection{General Architecture of the P4-MNA Implementation}
\label{sec:general_implementation}
A node running the P4-MNA implementation supports basic \acs{MPLS} functionality, i.e., \acs{MPLS} label pushing, popping, and swapping.
In addition, P4-MNA nodes parse and process network actions with the \acs{MNA} encoding according to the \acs{MPLS} stack structure described in \sect{placement}.
The implementation uses placeholders for the \acs{bSPL} value of the \acs{NAS} indicator and opcodes as they are yet to be assigned by IANA.
For simplicity, the implementation comprises \acs{HBH}-scoped \acs{NAS} and select-scoped \acs{NAS}, but not the \acs{I2E}-scoped \acs{NAS}.
However, the \acs{I2E}-scoped \acs{NAS} can be added using the same mechanisms described in the following.

The source code, including the data plane program in P4, and the control plane program in Rust  leveraging the rbfrt library~\cite{ZiFl25} is publicly available on GitHub~\cite{p4-mna-git}.
Further, we provide a Wireshark dissector for visualizing \acs{MNA} traffic, and a Python library to build MPLS stacks with \acs{MNA} \acp{LSE} in the GitHub repository.

\subsection{Parsing of the MPLS Stack}
\label{sec:mna_parsing}

\label{sec:impl_complexity}
The header encoding introduced with the \acs{MNA} framework in \sect{header} brings a lot of complexity to parsing because of different encodings and a dynamic structure.
In \acs{P4}, the parser is implemented as a simple finite-state machine.
On hardware targets such as the Intel Tofino™ 2, the parser is limited in its number of states and transitions due to hardware constraints.
A dynamic structure of different encodings in a header stack increases complexity by the number of states and transitions required for parsing.
The implementation of P4-MNA therefore facilitates the parsing of network actions by parsing all Format D \acp{LSE}, i.e., \acs{AD} \acp{LSE}, into the encoding of a Format C \acs{LSE} based on the length from the \acs{NASL} field.
This reduces the complexity of parsing the \acs{MNA} stack while preserving most of the semantics for different encodings.
The semantics between Format C and Format D \acp{LSE} are restored later in the control block using \acp{MAT}.
This is further explained in \sect{mna_processing}.

An example \acs{NAS} containing a Format B and multiple Format C and D \acp{LSE} with its parsed internal representation is shown in \fig{pdfs/ihle20}.

\figeps[\columnwidth]{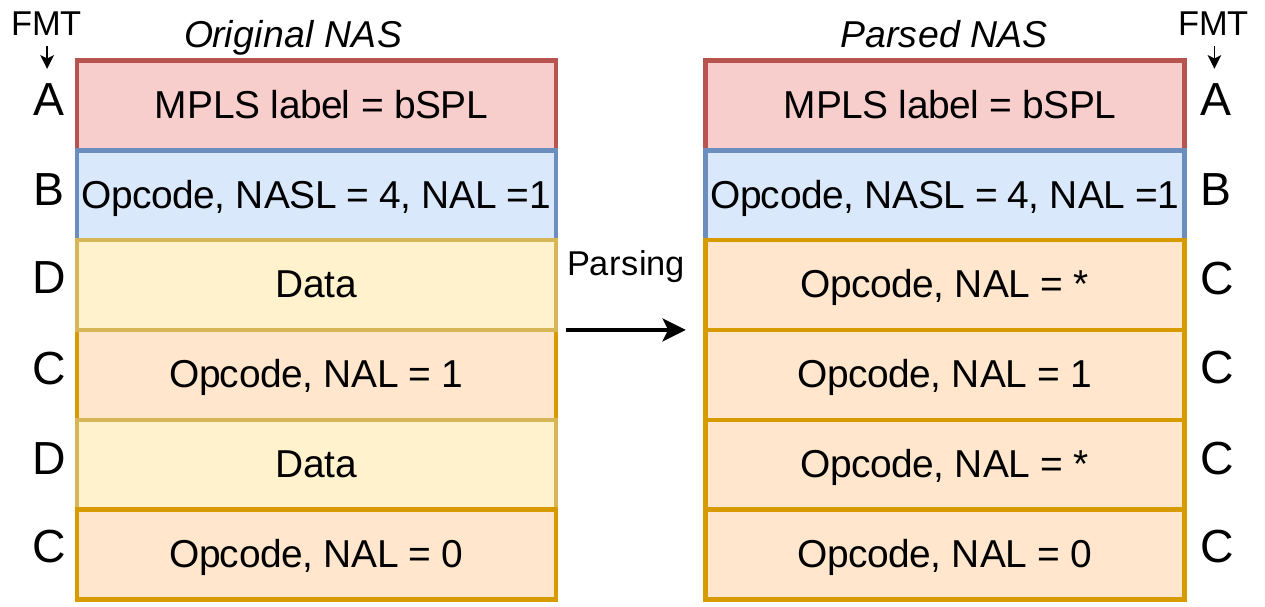}{All Format D \acp{LSE} are parsed in the encoding of a Format C \acs{LSE} to simplify the parser. The different encodings are later distinguished in the control blocks to restore the semantics of the Format D \acp{LSE}.}

In the example, the length of the \acs{NAS} is indicated by the value $4$ in the \acs{NASL} field in the Format B \acs{LSE}.
In P4-MNA, the example \acs{NAS} is parsed into one Format A and B \acs{LSE}, and four Format C \acp{LSE} according to the \acs{NASL} field.
The \acs{NAL} fields, i.e., the number of \acs{AD} \acp{LSE} per network action, are not relevant during parsing in P4-MNA.

\subsection{The P4-MNA Processing Pipeline}
\label{sec:mna_processing}
\revieweri{The processing pipeline of P4-MNA executes network actions through the application of \acp{MAT}.
For each of the 16 \acp{LSE} in a \ac{NAS}, one \acs{MAT} exists in the pipeline.
The \acp{MAT} share the same structure illustrated in \fig{pdfs/ihle21} but match on different indices in the array of parsed \acp{LSE}.}

\figeps[0.6\columnwidth]{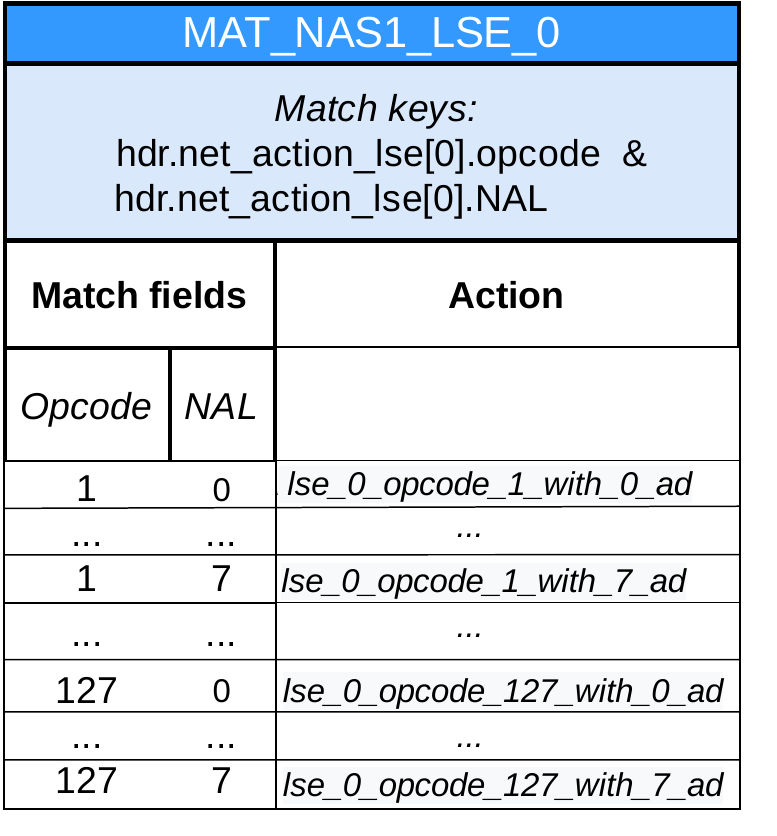}{A \acs{MAT} for a network action matches on the opcode and the \acs{NAL} field.}

In \fig{pdfs/ihle21}, a \acs{MAT} that executes network actions located in the first \acs{LSE}, i.e., at index $0$, of a \acs{NAS} is shown.
Such a \acs{MAT} matches on the opcode and the \acs{NAL} field of the Format B or Format C \acs{LSE}.
The \acs{MAT} has an action list with up to eight actions per opcode.
Only one of those actions is matched and applied per packet.
Each action accesses a different number of \acp{LSE} that follow the network action, i.e., \acs{AD} \acp{LSE}.
The number of accessed \acp{LSE} depends on the \acs{NAL} field in the matched network action.

The ingress control block of the P4-MNA pipeline applies the \acp{MAT} to a \acs{NAS}.
The processing pipeline of P4-MNA is shown in \fig{pdfs/ihle22}.

\figeps[0.9\columnwidth]{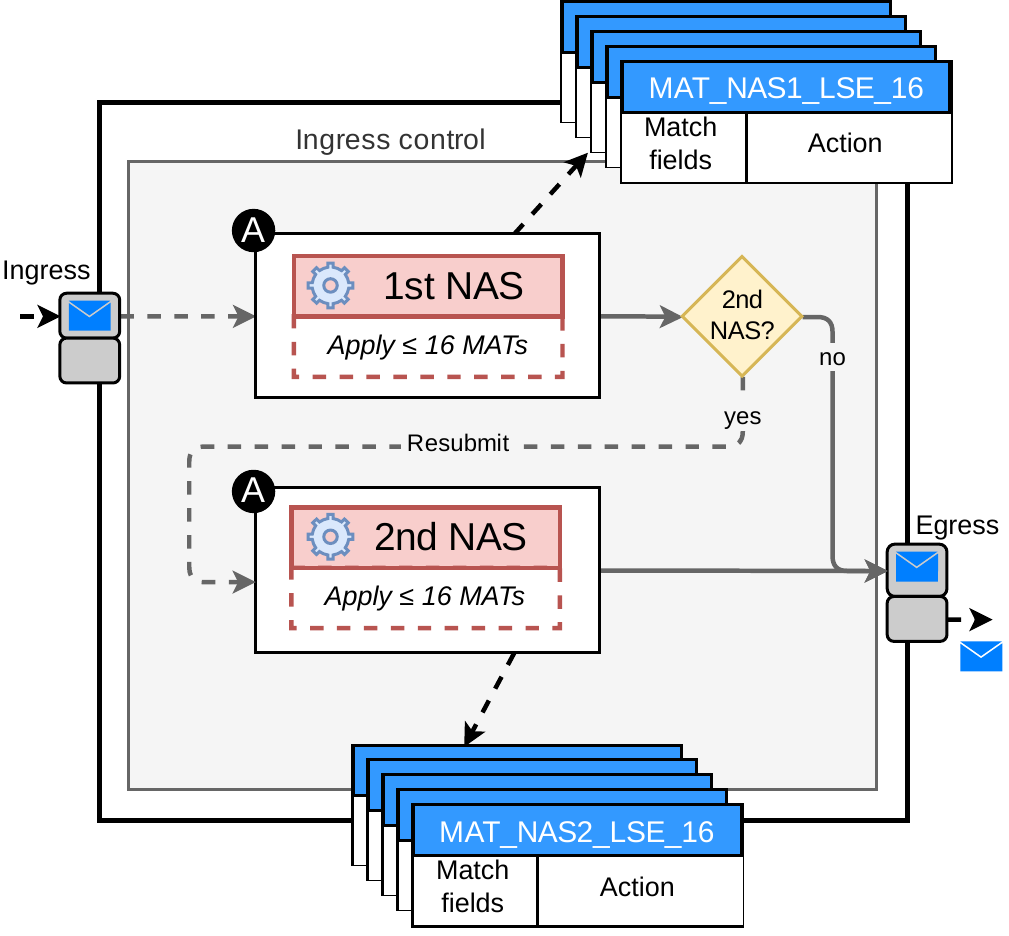}{The pipeline of P4-MNA can process up to 16 network actions in a single pipeline iteration. If more network actions are present, the packet is resubmitted.}

\revieweri{The P4-MNA pipeline can hold 16 \acp{MAT} per \acs{NAS} to match network actions.}
Therefore, if the packet contains multiple \acs{NAS}, it is resubmitted as only 16 of those \acp{MAT} fit into one pipeline iteration.
\revieweri{The processing of a NAS is referred to as \ballnumber{A} in \fig{pdfs/ihle22} and its operating principle is expanded in \fig{pdfs/ihle23}.}
During parsing, all Format D \acp{LSE} were parsed as Format C \acp{LSE} to facilitate the parsing of the complex \acs{MNA} header structure as described in \sect{impl_complexity}.
However, by doing this, the knowledge about the semantics of an \acs{LSE} is lost, i.e., whether the parsed \acs{LSE} is an \acs{AD} \acs{LSE}, or a network action \acs{LSE}.
To restore this information, a matched action accesses a number of \acp{LSE} that follow the network action according to the \acs{NAL} field and interprets those as \acs{AD} \acp{LSE}.
\revieweri{This mechanism is described in \fig{pdfs/ihle23} and expands block \ballnumber{A} from \fig{pdfs/ihle22}.}

\figeps[\columnwidth]{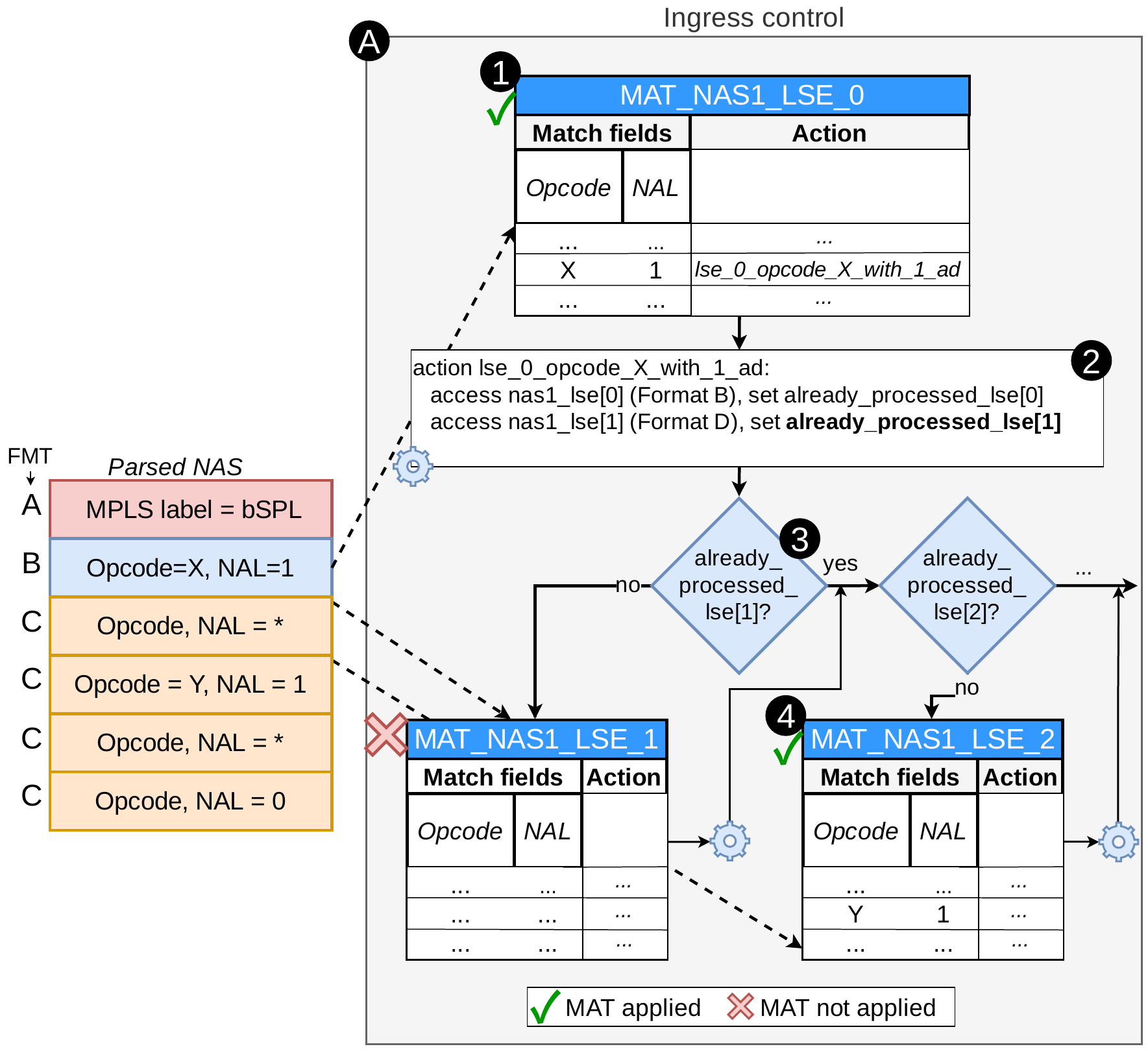}{A network action \acs{MAT} is applied to the first \acs{LSE} of the parsed \acs{NAS}. An entry is matched according to the opcode, and the \acs{NAL} field.
All accessed \acp{LSE} are marked as \texttt{already\_processed} and are not considered further in processing.
}

In \fig{pdfs/ihle23}, the control flow for the application of network action \acp{MAT} is shown with an exemplary \acs{NAS}.
The example \acs{NAS} corresponds to the parsed example from \fig{pdfs/ihle20}, i.e., all Format D \acp{LSE} are parsed as Format C \acp{LSE}.
In step \ballnumber{1}, the first \acs{MAT} is applied and matches on the first network action in the \acs{NAS}.
This network action has the opcode $X$ and one \acs{AD} \acs{LSE}, i.e., the \acs{NAL} field has the value $1$.
Therefore, the corresponding action with opcode $X$ and one \acs{AD} \acs{LSE} is executed in step \ballnumber{2}.
The actual logic of the network action is applied in this step.
The action accesses the \acs{LSE} indices $0$ and $1$, i.e., the network action LSE itself, and the succeeding \acs{AD} \acs{LSE}.
In the executed action, the first \acs{LSE} index is treated as Format B LSE\footnote{The first \acs{MAT} matches on the Format B \acs{LSE}, all other \acp{MAT} match on Format C \acp{LSE}.}.
All other accessed \acp{LSE} are treated as Format D \acp{LSE}.
Therefore, the semantics between the Format B/C and Format D encoding that were lost during parsing are restored in this step.
Further, in step \ballnumber{2}, the action marks both \acp{LSE} as \texttt{already\_processed} in the metadata.
\revieweri{Next, in step \ballnumber{3}, the control block checks if the succeeding \acs{LSE} was marked as \texttt{already\_processed} by a previous network action.
As the LSE at index $1$ in the example stack in \fig{pdfs/ihle23} was already accessed by the previous network action, the \acs{MAT} for this \acs{LSE} is not applied.
Consequently, the next \acs{LSE} index, i.e., index $2$, is checked whether it is marked as \texttt{already\_processed}.
As this is not the case, the \acs{MAT} for LSE index $2$ is applied in step \ballnumber{4} and the action with opcode $Y$ is executed.
This process repeats for all 16 \acp{MAT} in the pipeline.}

\subsection{Implemented Network Actions}
In this section we describe two implemented network actions, namely the performance measurement using the \acf{AMM} and bandwidth reservation with network slicing.

\subsubsection{Network Action for Performance Measurement Using AMM} 
\label{sec:amm_impl}
As a first example network action, we implement link-specific packet loss measurement using the \acf{AMM} as outlined in \sect{amm}. 
This implementation allows \acp{LSR} to track and report packet counters per flow enabling link-specific performance measurement.

To support AMM, a P4-MNA \acs{LSR} maintains two registers for each flow which count packets for the two colors used in AMM.
When an MPLS packet carrying a \acs{NAS} with the AMM network action is received, the LSR extracts the flow ID from the packet header and increments the corresponding color counter in the register.
If the color of two consecutive packets of the same flow differs, indicating a color change event, the \acs{LSR} generates a digest message containing the packet counter value, the flow id, and a timestamp.
This digest message is sent to the control plane which correlates the data received from all LSRs in the network.
By collating the timestamps included in the digest messages, the control plane computes the packet loss for each flow per hop.
This network action is evaluated in \sect{amm_eval}.

Time synchronization between LSRs is critical to ensure accurate correlation of the counters.
For practical deployments, external synchronization protocols such as the Network Time Protocol (NTP) or the Precision Time Protocol (PTP) are necessary to maintain consistent timing across distributed nodes.

\subsubsection{Network Action for Network Slicing} 
\label{sec:impl_nrp}
\revieweri{As a second example network action, we implement bandwidth reservation for network slices as outlined in \sect{network_slicing}.
The purpose of the MNA framework in network slicing is to carry the \acf{NRP} selector in each packet.
The \acs{NRP} selector identifies the network slice a packet belongs to.
The \acs{LSR} performs traffic shaping, e.g., bandwidth reservation, based on orchestrated \acp{NRP}.
The orchestration of \acp{NRP} is handled by a management and orchestration (MANO) framework and is considered out of scope for this work.}

\revieweri{
On processing the network slicing action in a P4-MNA \acs{LSR}, the NRP selector is extracted from the packet.
Next, the packet is matched in an additional \acs{MAT} that performs bandwidth metering using the meter extern of the Intel Tofino\texttrademark\ based on the extracted \ac{NRP} selector.
The reserved bandwidth per \acs{NRP} is configured by the control plane.
The bandwidth is metered according to the configured bandwidth reservation per network slice, i.e., traffic that exceeds the configured rate is dropped.
Other traffic, i.e., traffic that is not part of an \acs{NRP} is metered with the remaining available bandwidth.
This way, the bandwidth reservation for every \acs{NRP} is enforced through the NRP selector indication in the MNA network action.
This network action is evaluated in \sect{eval_nrp}.
}
\section{Evaluation}
\label{sec:evaluation}
\revieweri{In this section, we first analyze the scalability of P4-MNA in terms of the number of implementable network actions and the complexity of the network actions.
We then evaluate the impact of the number of network actions on the processing delay and forwarding rate, showing the ability of P4-MNA to process network actions with the MNA encoding.
Finally, we evaluate the two implemented use cases: link-specific packet loss measurement using \ac{AMM} and bandwidth reservation using network slicing.
The use case evaluations demonstrate the ability of P4-MNA to implement custom network actions using the available resources on the Intel Tofino\texttrademark\ 2 ASIC.}

\subsection{Scalability Analysis of P4-MNA}
\label{sec:scalability_eval}
The IETF \acs{MNA} encoding allows 128 different opcodes.
The P4-MNA pipeline described in \sect{mna_processing} requires up to eight entries in a \acs{MAT} and up to eight actions for one opcode, i.e., opcode $A$ with $0$ AD \acs{LSE}, opcode $A$ with $1$ \acs{AD} \acp{LSE} and so on.
In total, $128 \cdot 8 = 1024$ entries are needed to cover all combinations of opcodes with \acs{AD} \acp{LSE} in one \acs{MAT}.
A network action \acs{MAT} in P4-MNA has a size of 4096 table entries and can therefore hold all possible combinations.
However, when implementing a network action, additional resources such as registers for the \acs{AMM} network action, and meters for the network slicing action must be considered.

The complexity of network actions that can be implemented is limited to the functionality of the Intel Tofino™ 2 ASIC and the P4 language. 
Simple network actions such as the network slicing and the \acs{AMM} network action are implementable.
However, more sophisticated features, such as cryptographic functions, are not supported on the Intel Tofino™ 2.
More hardware, e.g., a cryptographic extern offloaded to an external server or a smartNIC, is required to implement such functionality.

\reviewerii{
More complex network actions may require more instructions that may not fit in the pipeline.
In this case, recirculation is required.
In a recirculation, a packet is looped back to an ingress port where it is processed again, i.e., the required instructions are applied.
An advantage of this is the implementation of more complex functions such as shown by~\cite{MeLi21}.
However, recirculation comes at the cost of bandwidth.
To avoid this, simple network actions should be preferred.
The network actions currently proposed by the WG are simple enough to not require recirculation.
However, a combination of many simple network actions may also exceed the packet processing resources within a single packet processing cycle and require recirculation.
}

\subsection{Impact of Number of Network Actions}
\revieweri{In this section, we evaluate the impact of the number of network actions on the processing delay and forwarding rate.
First, we describe the testbed and methodology.
Then, we present our results.}

\subsubsection{Testbed and Methodology}
\label{sec:testbed_example}
The P4-MNA implementation operates at line rate of \SI{400}{\gbps} per port.
We evaluate network action processing on a path of three \acp{LSR} emulated by ingress ports on an Intel Tofino™ 2 switch.
The ingress and egress \acp{LER} are emulated by a second Intel Tofino™ 2 switch running the P4TG traffic generator~\cite{p4tg}.
\revieweri{We extended P4TG to enable \SI{400}{\gbps} MNA traffic generation~\cite{IhZi24}.}
The testbed setup for this evaluation is shown in \fig{pdfs/ihle24}.

\figeps[0.9\columnwidth]{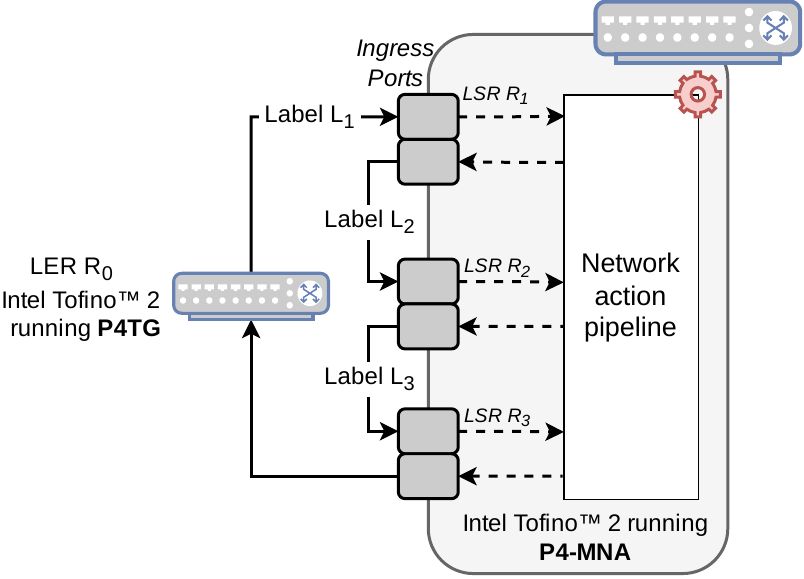}{The testbed consists of three \acp{LSR} represented as ingress ports on the P4-MNA switch. Packets containing the network actions are sent from P4TG to an ingress port of the switch. Here, network actions are processed and the packets are sent to another ingress port.}

In \fig{pdfs/ihle24}, packets are sent to an ingress port of the Intel Tofino™ 2 switch running the P4-MNA implementation.
Each ingress port of the P4-MNA switch represents an \acs{LSR} processing network actions.
After processing the network actions in the stack, \acp{LSR} pop the top-of-stack label and forward packets accordingly.
For label $L_1$, $L_2$, and $L_3$ packets are sent to a different ingress port on the same Intel Tofino™ 2 switch.
\acs{LSR} $R_3$ forwards packets back to the traffic generator P4TG for traffic analysis and measurement.
In the testbed setup, all LSRs operate on the same physical device and share a common time reference.

\revieweri{For all experiments, \ac{CBR} traffic is generated for approximately 60 seconds by P4TG.
The smallest frame size the traffic generator P4TG can generate to achieve \SI{400}{\gbps} is \SI{128}{\byte}.
Wide-area networks are dominated by Ethernet traffic which typically uses a default MTU size of \SI{1500}{\byte}~\cite{ChYo23}.
Therefore, we tested two packet sizes: \SI{128}{\byte} and \SI{1500}{\byte}.
For both tested packet sizes, the results were nearly identical with only a marginal reduction in latency for \SI{128}{\byte} packets, i.e., in the order of nanoseconds, resulting from the smaller packet size.
For clarity, we focus on the \SI{1500}{\byte} results as larger packet sizes are more representative of typical high-throughput network traffic.
Each experiment in the evaluation is repeated ten times and confidence intervals with a confidence level of \SI{99}{\percent} are calculated.}

\subsubsection{Performance Results}
\revieweri{In this evaluation, we test four different MPLS stacks and measure the RTT and packet loss to measure their impact  on the processing delay and forwarding rate.
\fig{pdfs/ihle25a_full} shows the MPLS stacks used in the experiments containing network actions and forwarding labels to reach the nodes $R_1$, $R_2$, and $R_3$ in a \acs{SR}-MPLS fashion.}

\foursubfigepsScale{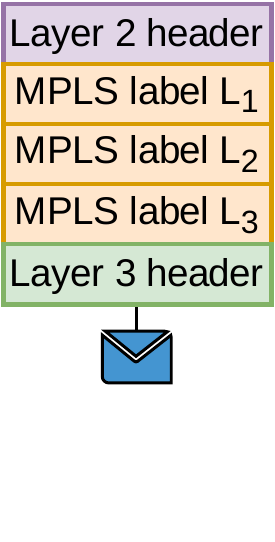}{E1.}{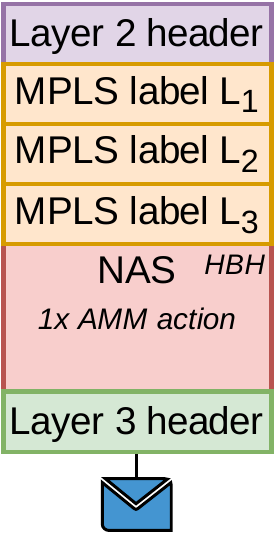}{E2.}{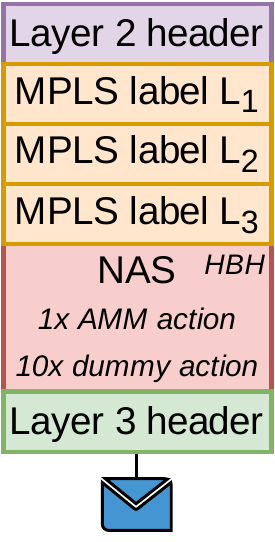}{E3.}{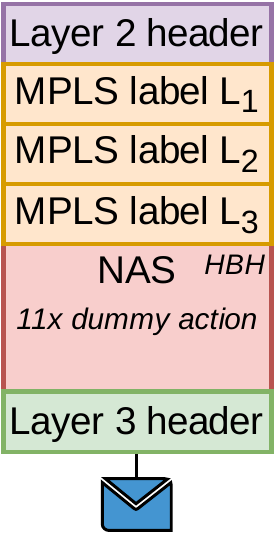}{E4.}{The \acs{MPLS} stacks used in the evaluation for experiments E1 -- E4.}


\revieweri{In the first experiment \textit{E1}, the MPLS stack contains only forwarding labels, i.e., labels $L_1$, $L_2$, and $L_3$ to validate the MPLS forwarding capability.
In the second experiment \textit{E2}, we add the \acs{AMM} network action from \sect{amm} in an \acs{HBH}-scoped \acs{NAS} to the MPLS stack.}
\revieweri{In the third experiment \textit{E3}, we add ten additional network actions to the HBH-scoped \acs{NAS} to see if multiple network actions have an impact on the processing delay.
Ten additional network actions is the maximum the traffic generator P4TG supports.
These are dummy network actions that write arbitrary data into their AD fields upon processing.
The dummy actions constitute a stress test to evaluate the system’s ability to process multiple network actions.
In the fourth experiment \textit{E4}, the MPLS stack contains an HBH-scoped NAS with eleven dummy network actions.
In this experiment, the AMM network action is not contained to see if the processing of the more sophisticated AMM network action impacts processing delay.
}

\revieweri{
This evaluation has two goals.
First, we want to verify if P4-MNA is capable of achieving line rate processing of network actions.
Second, we analyze the impact on the processing delay resulting from network actions in the MPLS stack.
We measure the packet loss PL and the RTT of the returned traffic using the measurement capabilities of the traffic generator P4TG.
The mean results of the RTT and packet loss measurement are shown in \tabl{results}.
For clarity, we omit confidence intervals as their width was less than \SI{0.004}{\percent} of the measured average for a confidence level of \SI{99}{\percent}.}

\begin{table}[htb!]
\caption{Measured packet loss $\overline{PL}$ and $\overline{RTT}$ for the experiments E1 -- E4.}
\begin{tabularx}{\columnwidth}{|X|X|X|X|X|X|X|}
\hline
                         & \textbf{E1} & \textbf{E2} & \textbf{E3} & \textbf{E4} \\ \hline \hline
$\overline{PL}$              & 0               & 0             & 0    & 0              \\ \hline
$\overline{RTT}$            & \SI{4.75}{\micro\second}    & \SI{4.79}{\micro\second}  & \SI{4.79}{\micro\second}& \SI{4.79}{\micro\second} \\ \hline
    \end{tabularx}
    \label{tab:results}
\end{table}

\revieweri{\tabl{results} shows that no packet loss was recorded, i.e., P4-MNA achieves line rate forwarding of \SI{400}{\gbps} with MPLS traffic in the experiments E1 -- E4.
Adding network actions increases the RTT marginally by $\approx$ \SI{40}{\nano\second} from \SI{4.75} {\micro\second} in E1 to \SI{4.79}{\micro\second} in E2 -- E4.
A packet is processed once by each \acs{LSR}, i.e., three times in total\footnote{The per-packet processing delay is constant on the Intel Tofino\texttrademark.}.
The latency added by one \acs{LSR} in E2 -- E4 is therefore $\frac{40\ ns}{3} = 13.3\ ns$.}

\revieweri{
The increased transmission delay resulting from the larger frame size due to additional network actions is negligible, i.e. less than \SI{1}{\nano\second}.
According to a latency profiling study of the Tofino\texttrademark\ P4 programmable ASIC-based hardware by Franco \textit{et al.}, the number of applied tables and the number of parser states increase the latency in the order of nanoseconds~\cite{FrZa24}.
The number of applied tables differs in E2 and E3 because more network actions exist in the MPLS stack, but both experiments result in the same latency.
Therefore, the number of applied tables has no measurable impact on the processing delay in P4-MNA.
In E2 and E3, the AMM network action is applied, which performs more sophisticated operations compared to the dummy actions such as accessing registers.
However, there is no increase in RTT compared to E4 where only dummy network actions are applied.
Therefore, the complexity of network actions does not affect the processing delay in P4-MNA.
The difference between E1 and E2 -- E4 is the additional parsing state that extracts network actions.
We therefore attribute the increase in RTT to the additional parsing state.
However, the observed RTT increase of \SI{13.3}{\nano\second} per hop is negligible.}

\revieweri{
Although the current evaluation is limited to eleven network actions due to traffic generation constraints, the P4-MNA implementation supports up to 32 network actions.
In future work, we plan to extend the P4TG traffic generator to generate larger MPLS stacks to further evaluate the P4-MNA implementation.}

\subsection{Use Case 1: Alternate Marking Method (AMM)}
\revieweri{In this section we describe the evaluation of the first implemented use case, i.e., link-specific packet loss measurement using the alternate marking method (AMM).
For this evaluation, the testbed and methodology described in \sect{testbed_example} are used.
First, we describe the configuration and validation of end-to-end packet loss in the testbed.
Then, we use the AMM network action to calculate link-specific packet loss.}

\subsubsection{Validation of End-to-End Packet Loss}
\label{sec:eval_drop}
\revieweri{
In experiment \textit{E5}, the \acp{LSR} $R_1$, $R_2$, and $R_3$ are configured to drop packets probabilistically to emulate packet loss on a link.
In this evaluation, we verify the configured end-to-end packet loss.
This experiment serves as a baseline for evaluating the link-specific packet loss measurement using the implemented AMM network action in experiment E6.
The probability $p_{i,i+1}^{drop}$ denotes the packet drop probability on the link from node $R_i$ to $R_{i+1}$.
We configure the probabilities $p_{0,1}^{drop} = 0.1$, $p_{1,2}^{drop} = 0.2$, and $p_{2,3}^{drop} = 0.3$.
The packets are dropped according to the configured probability in the ingress of node $R_{i+1}$ before any packet processing is applied.
The end-to-end packet loss is measured at the traffic generator P4TG.
}

\revieweri{For experiment E5, i.e., with probabilistic packet drop, the measured packet loss at P4TG is at \SI{49.6}{\percent}.
The expected end-to-end drop probability $p_{0,3}^{drop}$ of a frame being dropped by one of the \acp{LSR} is given in \equa{drop_prob}.}
\begin{equation}
    p_{0,3}^{drop} = 1- \prod_{i\in\{0,1,2\}}(1-p_{i,i+1}^{drop})
    \label{eq:drop_prob}
\end{equation}

\revieweri{For the configured drop probabilities the expected end-to-end drop probability of a frame is $p_{0,3}^{drop} = 0.496$.
The measured packet loss of \SI{49.6}{\percent} matches exactly with the expected probability, validating the configuration.}

\subsubsection{Validation of Link-Specific Packet Loss}
\label{sec:amm_eval}
\revieweri{In experiment \textit{E6}, we evaluate the link-specific packet loss measurement of the implemented AMM network action.
The \acp{LSR} are configured to drop packets on ingress according to experiment E5 in \sect{eval_drop}.
We calculate the link-specific packet loss resulting from the reported packet counters of the AMM network action and compare them to the configured packet loss.
In this experiment, the color in the generated AMM network action is alternated ten times approximately every five seconds at a rate of \SI{400}{\gbps}.
Every \acs{LSR} exports its packet counter value to the control plane when the color changes.
For evaluation, the control plane uses the last counter value received to calculate the packet loss of each link.
}

\revieweri{
For a link from an LSR $R_i$ to its successor LSR $R_{i+1}$, the link-specific packet loss $\Delta_{i,i+1}$ is calculated in the control plane based on the reported counter values triggered by the AMM network action.
The value $N_i^a$ denotes the counter value of \acs{LSR} $R_i$ and color $a$, i.e., the number of received packets of color $a$ at LSR $R_i$.
\equa{packet_loss0} calculates the link-specific packet loss between \ac{LSR} $R_i$ and $R_{i+1}$ based on both color counter values.
This equation can be further simplified to \equa{packet_loss}.}

\begin{align}
    \label{eq:packet_loss0}
    \Delta_{i,i+1} &= (N_i^a-N_{i+1}^a)+(N_i^b-N_{i+1}^b)\\
    &= (N_i^a+N_{i}^b)-(N_{i+1}^a+N_{i+1}^b)\\
    &= N_i^{total}-N_{i+1}^{total}
    \label{eq:packet_loss}
\end{align}

\revieweri{
\equa{packet_loss} measures the link-specific packet loss $\Delta_{i,i+1}$ by correlating the counter values of two consecutive \acp{LSR} $R_i$ and $R_{i+1}$ regardless of packet color.
In the evaluation, we can use \equa{packet_loss} to calculate the link-specific packet loss.
In general, however, both colors are required to have a synchronized export trigger as described in \sect{amm}.}

\revieweri{
In the evaluation testbed, LSR $R_1$ does not have a preceding \acs{LSR}.
To calculate the packet loss $\Delta_{0,1}$ on the link from the traffic generator to LSR $R_1$, the total number of sent packets is known to the control plane.}

\revieweri{The formula for the calculated link-specific packet loss rate $\hat{p}_{i,i+1}^{drop}$ on a link from LSR $R_i$ to $R_{i+1}$ is shown in \equa{loss_rate}.}

\begin{equation}
    \hat{p}_{i,i+1}^{drop} = \frac{\Delta_{i,i+1}}{N_i^{total}}
    \label{eq:loss_rate}
\end{equation}

\revieweri{The mean results from experiment E6 are shown in \tabl{amm_results}.
The width of the confidence intervals was less than \SI{0.005}{\percent} of the measured average and therefore, they are not shown.}

\begin{table}[htb!]
	\caption{Mean counter values and calculated packet loss per link with the \acs{AMM} network action.}
    \begin{tabularx}{\columnwidth}{|X|X|X|X|X|}
\hline
\textbf{}                                                                 & \textbf{Generated packets} & \textbf{LSR $R_1$} & \textbf{LSR $R_2$} & \textbf{LSR $R_3$} \\ \hline \hline
\textbf{\#Received packets of color \textit{a} $N_{i+1}^a$}                                                          &                         & 860879414       & 688688859       & 482063758       \\ \hline
\textbf{\#Received packets of color \textit{b} $N_{i+1}^b$}                                                          &                         & 860953198       & 688751879       & 482136175       \\ \hline
\textbf{$N_{i+1}^{total}$}                                                            & 1913168832              & 1721832612      & 1377440738      & 964199933      \\ \hline
\textbf{Calc. loss with AMM $\Delta_{i,i+1}$} &            & 191336220       & 344391874       & 413240805       \\ \hline
\textbf{Calc. loss rate with AMM $\hat{p}_{i,{i+1}}^{drop}$}     &                                                 & 0.1000       & 0.2000       & 0.3000       \\ \hline
\textbf{Conf. loss $p_{i,{i+1}}^{drop}$}                                             &     & 0.1       & 0.2       & 0.3       \\ \hline
    \end{tabularx}
    \label{tab:amm_results}
\end{table}

\revieweri{
\tabl{amm_results} shows that the calculated link-specific packet loss rate $\hat{p}_{i,{i+1}}^{drop}$ matches the configured drop probabilities $p_{i,{i+1}}^{drop}$ for each LSR.
For \acs{LSR} $R_1$, \SI{10}{\percent}, for \acs{LSR} $R_2$ \SI{20}{\percent}, and for \acs{LSR} $R_3$ \SI{30}{\percent} packet loss was measured.
This matches the configured drop probability at each \acs{LSR} exactly.
Thus, the implemented AMM network action is capable of precisely measuring link-specific packet loss.}

\subsection{Use Case 2: Network Slicing}
\revieweri{In this section we describe the evaluation of the second implemented use case, i.e., bandwidth reservation with network slicing.
We apply the same methodology as explained in \sect{testbed_example} but use a different testbed.
First, we describe the testbed, and then, present our results.}

\subsubsection{Testbed}
\label{sec:testbed_nrp}
\revieweri{In experiment $E7$, we evaluate the implemented network slicing action for bandwidth reservation described in \sect{impl_nrp}.
We use the testbed as shown in \fig{pdfs/ihle26}.}

\figeps[0.75\columnwidth]{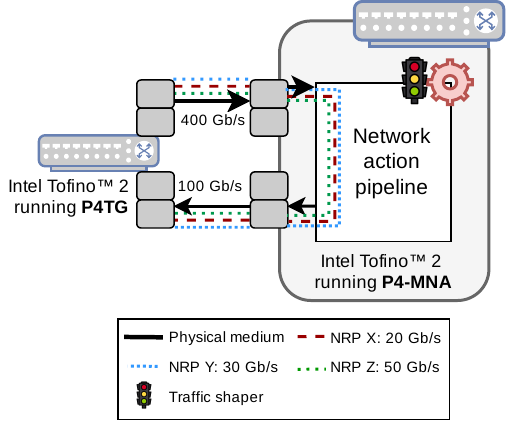}{Three \acp{NRP} are configured with bandwidth reservations. They are generated by P4TG and are sent via a 400 Gb/s link to the P4-MNA switch. The traffic is returned via a 100 Gb/s bottleneck link to P4TG for measurement.}

\revieweri{The testbed in \fig{pdfs/ihle26} contains two Intel Tofino\texttrademark\ 2 switches that are connected with a physical \SI{400}{\gbps} and a \SI{100}{\gbps} link.
The first Tofino\texttrademark\ runs the P4TG traffic generator and the second runs the P4-MNA implementation.
In the testbed, three \acp{NRP} are configured.
NRP X has a bandwidth reservation of \SI{20}{\gbps}, NRP Y of \SI{30}{\gbps}, and NRP Z of \SI{50}{\gbps}.
P4TG generates three streams containing a network action with the NRP selectors X, Y, and Z.
They are generated with the respective reserved bandwidth and are sent via the \SI{400}{\gbps} link to the P4-MNA switch.
The P4-MNA switch performs traffic shaping according to the configured bandwidth reservations of the \acp{NRP} and the extracted \acs{NRP} selector from the network action.
For that purpose, the control plane configures the meter extern in the data plane with the reserved bandwidth for each \acs{NRP}.
Then, the switch returns the traffic via the \SI{100}{\gbps} link to P4TG for measurement.
The \SI{100}{\gbps} link creates a bottleneck in the network that causes congestion in the switch if the capacity of the link is exceeded leading to packet loss.
We add a fourth stream generated on the \SI{400}{\gbps} link that exceeds the bottleneck link's capacity to force congestion on the egress link.
This stream is interference traffic and does not contain an NRP selector.
Therefore, no bandwidth is reserved for the interference traffic stream.
In total, \SI{200}{\gbps} of traffic is generated.}

\subsubsection{Validation of Bandwidth Reservations}
\label{sec:eval_nrp}
\revieweri{
First, we measure the packet loss per stream and the RTT without generating interference traffic, i.e., no congestion is caused on the link.
Next, we add interference traffic but do not apply the network slicing action, leading to congestion.
Finally, we apply the network slicing action.
The mean measured packet loss per stream for $E7$ is shown in \fig{pdfs/ihle27}.
The width of the confidence intervals was less than \SI{0.05}{\percent} of the measured average and therefore invisible.}

\figeps[\columnwidth]{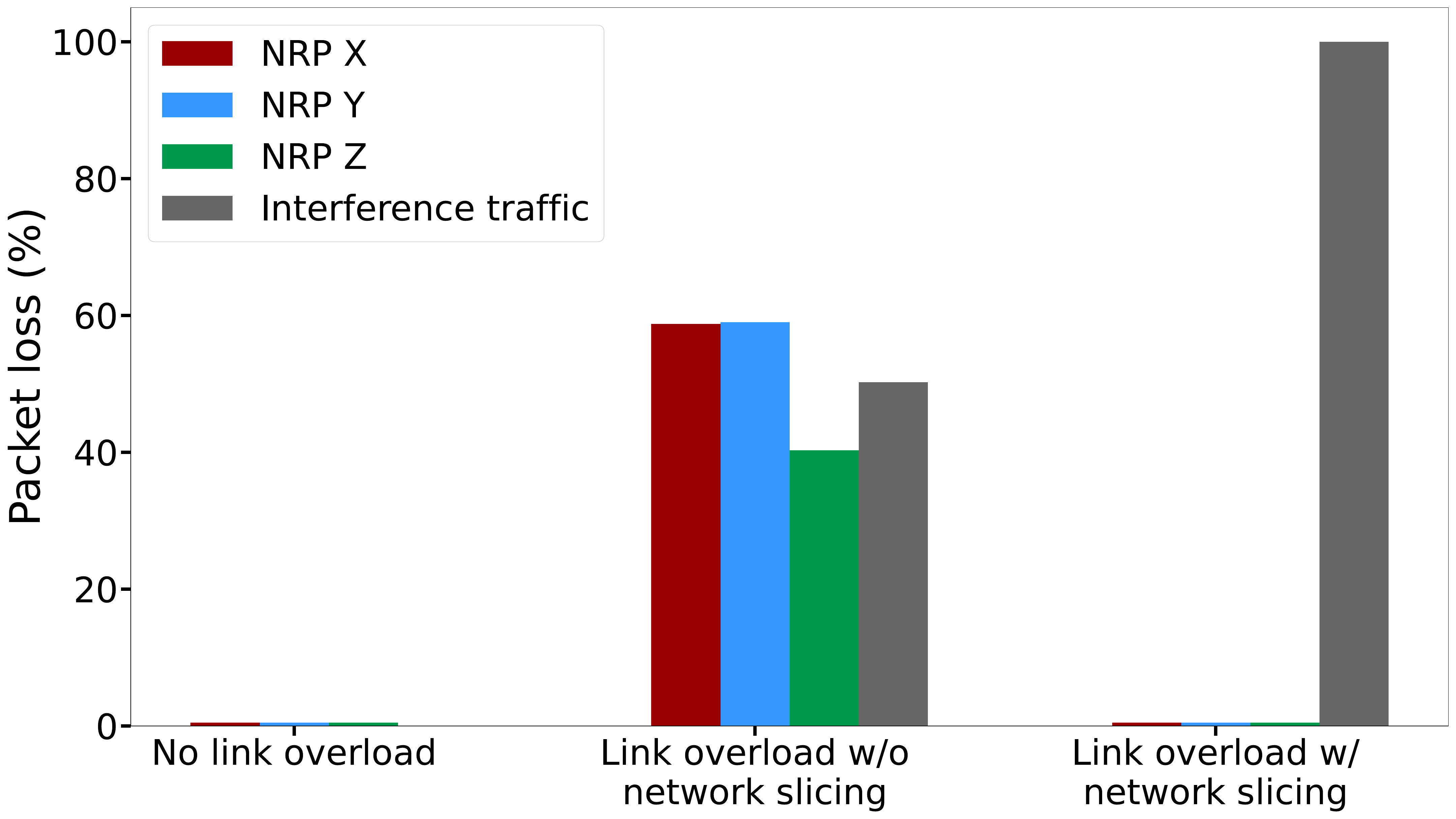}{Mean packet loss measured by P4TG with and without traffic shaping according to NRPs.}

\revieweri{In \fig{pdfs/ihle27}, no packet loss is visible if the link is not overloaded.
This is the baseline.
If the interference traffic is added and the \acp{NRP} are not enforced by the switch, all streams see a high packet loss between \SI{40}{\percent} and \SI{60}{\percent}\footnote{We observe different drop probabilities for the different flows although they travel jointly over the same link. The reason for this phenomenon is that the traffic is strictly \ac{CBR}. Therefore, we see periodic arrival behavior at the bottleneck link and packets of some flows get dropped more often than packets of other flows within a period due to combinatorial effects.}.
Therefore, the reserved bandwidth for the \acp{NRP} is not guaranteed.
If the \acp{NRP} are enforced by the switch, the interference traffic is dropped.
As a result, the egress link is not overloaded and the \acs{NRP} streams see no packet loss.
Their reserved bandwidth is guaranteed.}

\revieweri{This evaluation has shown that the MNA network slicing action can indicate \acp{NRP} in the MPLS stack to enforce network resource reservations.
The focus in this work lies on the indication of the NRP selector in a network action.
For practical use, a management and orchestrator (MANO) framework is required.
More information about network slicing with bandwidth guarantees in P4 switches can be found in~\cite{ChCh22}.}

\section{Constraints for Header Stacking in MNA}
\label{sec:eval_signaling}
\revieweri{In this section, we explain two constraints that apply to header stacking in \acs{MNA}.
First, we describe a constraint resulting from the available resources (RLD) in hardware.
Then, based on our implementation, we describe a constraint arising from the maximum \acs{NAS} sizes that need to be supported in the proposed \acs{MNA} framework.
As the latter constraint makes LSRs with small RLD not usable for MNA, we suggest that LSRs may support lower NAS sizes and to signal these values in addition to the RLD so that hardware with lower processing capabilities can be integrated into MNA domains.}

\subsection{Constraints for Header Stacking due to Readable Label Depth (RLD)}
\label{sec:constraint_rld}
\reviewerii{
In general, hardware can only parse limited bytes of a packet header.
This limitation results from the hardware resources available on the device such as memory.
Additionally, the complexity of a packet parser is limited to facilitate packet processing within a single packet cycle.
Header sizes that exceed the available resources for parsing cannot be processed in a single packet cycle and, therefore, are an obstacle for packet forwarding at line rate.
Thus, for practical deployment, the relevant information in the header must fit within the readable bytes.
In MNA, this translates to the so-called \acf{RLD}~\cite{ietf-mpls-mna-fwk-05} which indicates the number of \acp{LSE} a node can parse.
}

\revieweri{The \acs{RLD} is a critical parameter of \acp{LSR} when deploying \acs{MNA} in \acs{MPLS} networks.
In the current proposal, \acp{LSR} signal their \acs{RLD} to the ingress \acp{LER} using a routing protocol such as IS-IS~\cite{rfc9088} or OSPF~\cite{rfc9089}.
The ingress \acs{LER} must ensure that a \acs{NAS} determined for a node is within the \acs{RLD} when the packets reach that node.
This is done based on the signaled \acs{RLD} parameter.
An \acs{HBH}-scoped \acs{NAS} can be located at the bottom of stack.
However, if this \acs{HBH}-scoped \acs{NAS} is not within the RLD of an intermediate node, the ingress LER places copies of that \acs{NAS} within the \acs{RLD} to ensure that the NAS can be found by that node.}
%
\revieweri{
Only the topmost \acs{HBH} \acs{NAS} copy is processed by a node.
An example is given in \fig{pdfs/ihle28} and further described below.}

\figeps[\columnwidth]{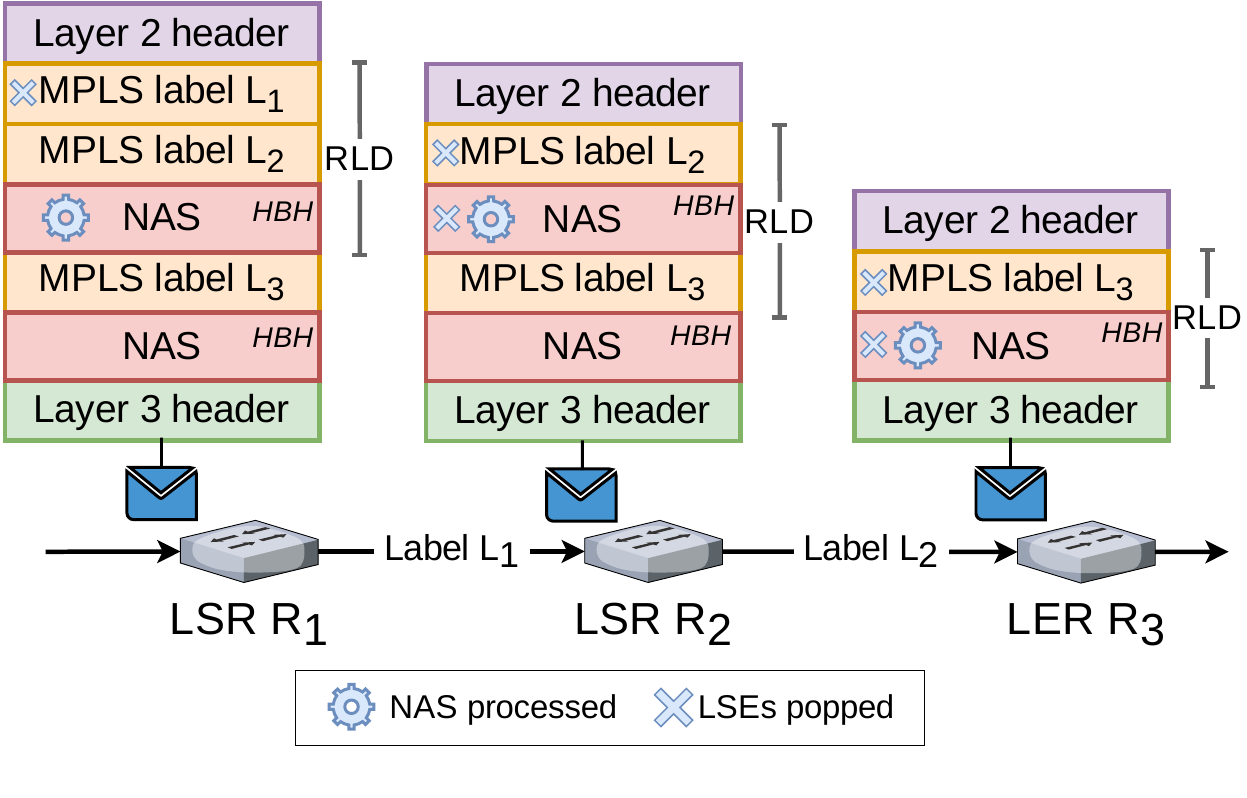}{An example \acs{MPLS} label stack with one \acs{HBH}-scoped \acs{NAS} in the \acs{RLD} for each node. As the \acs{RLD} of \acs{LSR} $R_1$ and $R_2$ is limited to 3 \acp{LSE}, a copy of the \acs{HBH}-scoped NAS is placed within the RLD of those \acp{LSR}.}

\revieweri{In \fig{pdfs/ihle28}, the nodes have an \acs{RLD} of three \acp{LSE}\footnote{For simplicity in the example, we treat a \acs{NAS} as a single \acs{LSE}. In reality, each \acs{NAS} may contain up to $17$ \acp{LSE}.}.
Placing a single \acs{HBH}-scoped \acs{NAS} at the bottom-of-stack is therefore not sufficient because it is not within the \acs{RLD} of \acs{LSR} $R_1$.
The ingress \acs{LER} must place the \acs{HBH}-scoped \acs{NAS} below the forwarding label $L_2$ and a copy of the \acs{HBH}-scoped \acs{NAS} below forwarding label $L_3$ to ensure that all nodes can process the \acs{NAS} in their \acs{RLD}.
$R_1$ parses all \acp{LSE} in its \acs{RLD} and processes the \acs{HBH}-scoped \acs{NAS} found below the forwarding label $L_2$.
$R_2$ processes the \acs{HBH}-scoped \acs{NAS} found below the forwarding label $L_2$ and pops that \acs{NAS} after processing.
Finally, $R_3$ processes the copy of the \acs{HBH}-scoped \acs{NAS} found below the forwarding label $L_3$.}

\subsection{Constraints for Header Stacking due to Maximum NAS Sizes}
\revieweri{Both select-scoped NAS and HBH-scoped NAS can contain up to 17 LSEs. 
Since a node may need to process both a select-scoped NAS and an HBH-scoped NAS, it requires two arrays of maximum NAS size to parse both NAS.
We denote these sizes as $maxLSEs^{select}_{NAS}$ and $maxLSEs^{HBH}_{NAS}$.
Since the HBH-scoped NAS may not immediately follow the select-scoped NAS but be located deeper in the stack, the LSEs in between are parsed into a third array which we call the in-between stack and which is $maxLSEs^{btwn}_{stack}$ LSEs large.
Thus, the RLD comprises one forwarding label, a maximum select-scoped NAS, the in-between stack, and a maximum HBH-scoped NAS.
As $maxLSEs^{select}_{NAS}$ and $maxLSEs^{HBH}_{NAS}$ are 17 LSEs according to the proposed standard~\cite{ietf-mpls-mna-hdr-04}, and the RLD is given by the hardware, the size of the in-between stack can be computed by \equa{btwn}.}

\begin{equation}
\begin{aligned}
    \text{maxLSEs}^{btwn}_{stack} = RLD 
    &- \text{maxLSEs}^{select}_{NAS}\\
    &\quad - \text{maxLSEs}^{HBH}_{NAS} - 1.\label{eq:btwn}
\end{aligned}
\end{equation}

\revieweri{This holds under the assumption that the hardware implementation cannot share the memory for parsing the select-scoped NAS, the HBH-scoped NAS, and the in-between stack, which is the case for P4-MNA.
The relation between the three arrays and the RLD is depicted on the left side in \fig{pdfs/ihle29}.
In case of P4-MNA, $RLD=51$, therefore, the in-between stack can contain up to 16 LSEs.
They contain any LSE between the two NAS, i.e., forwarding labels, special purpose labels, and network actions.
This seems a low value compared to a maximum NAS size.
Moreover, we conclude from \equa{btwn} that MNA-capable hardware must have an RLD of at least 35 LSEs, which may be problematic for some hardware platforms.}

\figeps[\columnwidth]{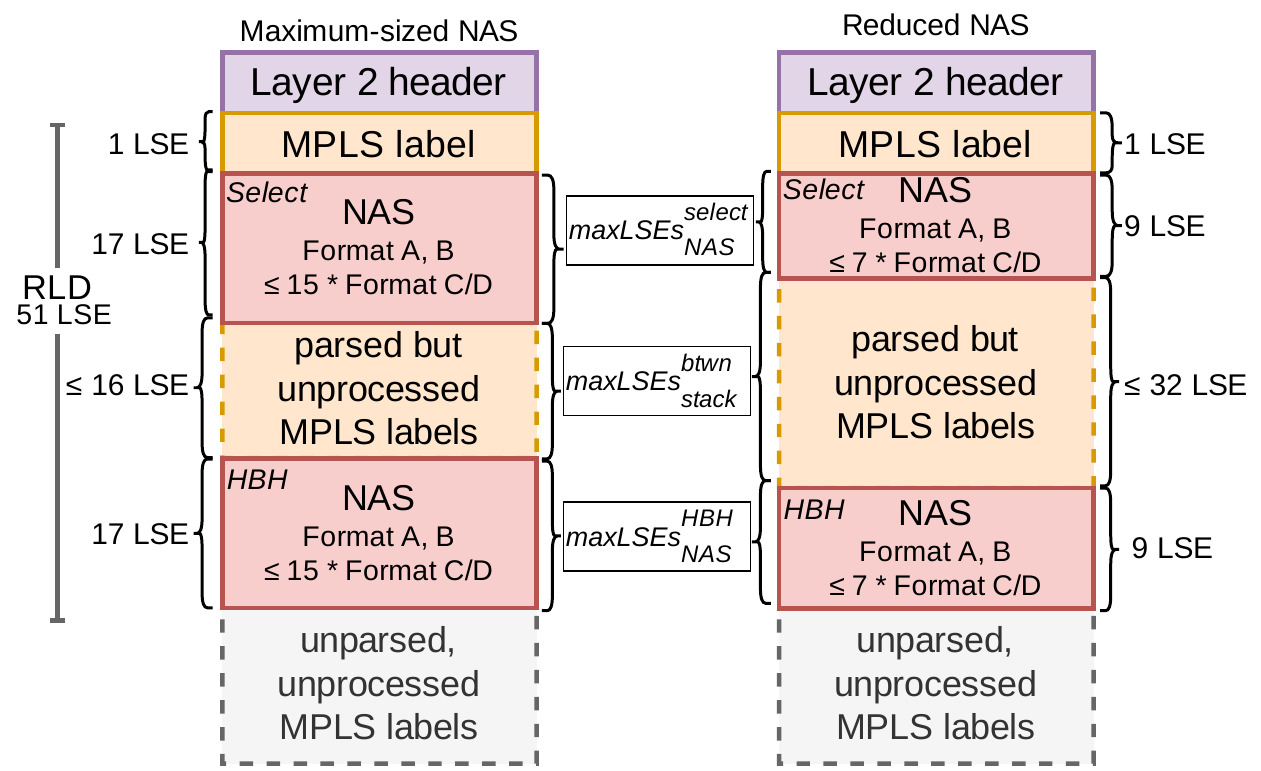}{The maximum size of a \acs{NAS} in the IETF proposal is $maxLSEs^{select}_{NAS} = maxLSEs^{HBH}_{NAS} = 17$.
P4-MNA has a RLD of 51 \acp{LSE}. This leaves 17 \acp{LSE} for other purposes.
If the parameters $maxLSEs^{select}_{NAS}$ and $maxLSEs^{HBH}_{NAS}$ are reduced, more \acp{LSE} are available for other purposes.}

\subsection{Signaling Maximum NAS Sizes}
\revieweri{The problems observed with a small in-between stack or a minimum RLD of 34 LSEs for MNA-capable hardware are due to the fact that a NAS can be up to 17 LSEs in size.
However, many network actions, or even a combination of many actions, do not require 17 \acp{LSE} in a \acs{NAS}.
We now propose that a NAS can be up to 17 LSEs in size, but nodes can also support smaller NAS.
To avoid receiving packets with larger NAS, we suggest that the node-specific maximum NAS sizes $maxLSEs^{select}_{NAS}$ and $maxLSEs^{HBH}_{NAS}$ are signaled to all ingress LERs, just like the RLD.
This allows hardware with a lower RLD to be used for MNA.}

\revieweri{An example is given on the right side of \fig{pdfs/ihle29}.
We choose $maxLSEs^{select}_{NAS}=maxLSEs^{HBH}_{NAS}=9$ for P4-MNA.
With an RLD of 51 LSEs this leaves $maxLSEs^{btwn}_{stack}=32$ LSEs for the in-between stack according to \equa{btwn}.
Moreover, there is no longer a minimum RLD for MNA-capable nodes as the maximum NAS sizes can be adjusted to fit within the RLD of the node.
However, in turn, the supported NAS sizes may be too small for some applications.}




\section{Challenges with Mutable Data in an \acs{ISD} Implementation}
\label{sec:challenges}
In this section, we explain the concept of mutable data and describe a challenge when using mutable data with \acf{ISD}.

Mutable data in \acs{MNA} refers to data in which the value in one of the data fields of a \acs{NAS} changes during packet forwarding.
An example of mutable data in \acs{MNA} is the collection of telemetry data along a path in the context of I\acs{OAM} in passport mode.
Here, each node on the path adds information to the \acs{NAS}, such as its node ID.
This information must not be discarded during packet forwarding, i.e., the \acs{NAS} containing this information must not be popped.

\revieweri{For backward compatibility, the range of use cases that can be implemented with \acs{ISD} is limited because mutable data is constrained.
This constraint results from the protocol design and not from the P4-MNA implementation.
The first 20 bits, i.e., the \acs{MPLS} label, are often hashed for \acs{ECMP} load balancing and therefore must not be altered in transit.
The number of \acp{LSE} used for hashing varies in implementations from one up to 16 \acs{MPLS} labels~\cite{juniper_hash, huawei_hash, cisco_hash}.
Entropy labels\cite{RFC6790} may be introduced to account for ECMP load balancing and to avoid the need to hash the \acs{MPLS} stack.
However, this does not provide full backward compatibility as legacy devices may not support entropy labels.}

\revieweri{A \acs{NAS} with the maximum possible number of mutable bits according to the MNA encoding is illustrated in \fig{pdfs/ihle30}.
The first 20 bits of an \acs{LSE} are hashed for \acs{ECMP} and must not be modified.
The remaining 12 bits are partially occupied by length fields and other flags, such as the bottom of stack bit.
Therefore, in a Format B \acs{LSE}, zero bits, in a Format C \acs{LSE} seven bits, and in a Format D \acs{LSE}, eleven bits are mutable.}

\figeps[\columnwidth]{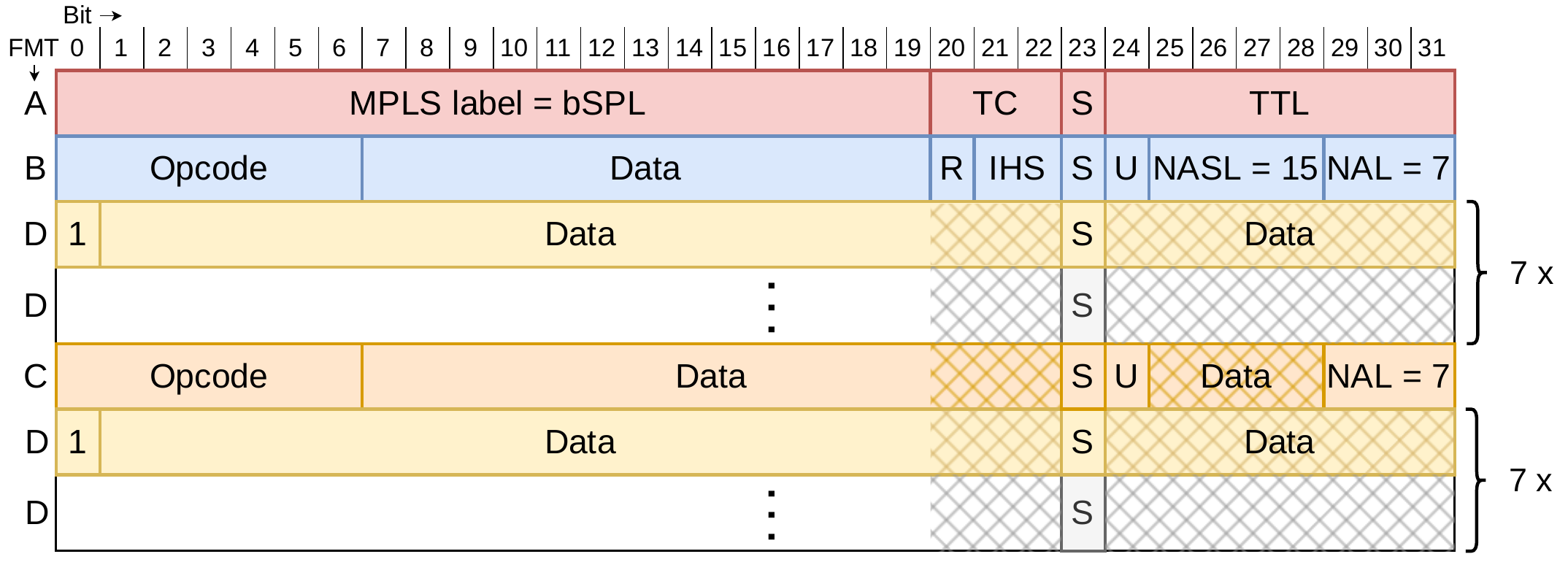}{\FI{A \acs{NAS} carrying the maximum number of mutable bits (hatched) consists of a Format B \acs{LSE}, followed by seven Format D \acp{LSE}, one Format C \acs{LSE}, and seven Format D \acp{LSE}.}}

\revieweri{In \fig{pdfs/ihle30}, a \acs{NAS} with $424$ data bits is shown.
However, only $161$ of those data bits are mutable.
The overall size of the \acs{NAS} is $544$ bits (\SI{68}{\byte}).
Therefore, only a fraction of the \acs{NAS}, i.e., $\approx\ 30\%$, can be used for mutable data.
Further, the number of \acs{AD} \acp{LSE} for a single network action is limited to seven.
Therefore, only $77$ mutable bits are available to a network action in the Format B encoding, and $84$ bits in the Format C encoding.}

The passport method described for IOAM in \sect{oam} collects information from \acp{LSR} along a path, e.g., ingress timestamps.
Here, each \acs{LSR} adds a timestamp to the packet.
Because collected timestamps are 32-bit wide\cite{rfc9197}, an in-stack network action can carry only two timestamp values as mutable \acs{ISD} for an entire \acs{LSP}.

If a network action requires a lot of mutable data, \acs{ISD} is inefficient because it wastes bits that cannot be mutated.
Alternatively, mutable data can be carried as \acf{PSD} where all bits are mutable~\cite{ietf-mpls-ps-mna-hdr-00}.
With \acs{PSD}, \acs{AD} is placed after the bottom-of-stack.
An implementation of the \acs{MNA} framework leveraging \acs{PSD} is out-of-scope for this work but will be explored in the future.

When using SR-MPLS, the top-of-stack forwarding label is popped per segment.
Therefore, a \acs{NAS} below the forwarding label is also popped.
These \acs{NAS} cannot contain mutable data because they do not reach the destination.
When SR-MPLS is used with in-stack mutable data, the mutable data must be at the bottom of the stack to avoid them being popped.
In this case, the \acs{RLD} of each \acs{LSR} must be large enough to parse the entire MPLS stack, i.e., the mutable data located in the \acs{NAS} at the bottom of the stack.

\section{Security Considerations}
\label{sec:security}
\reviewerii{
In this section we describe security risks and considerations arising from the use of the MNA framework.
Currently, there are three security risks that are considered in the MNA drafts~\cite{ietf-mpls-mna-fwk-05, ietf-mpls-mna-hdr-04}.
The first is link-level security, the second is MNA information originating from other domains, and the third is the syntax and semantics of network actions.}

\subsection{Link-Level Security}
\reviewerii{
The information contained in network actions is sensitive and may be exploited or manipulated for attacks.
Thus, MNA traffic should be protected from eavesdropping and manipulation.
However, the MPLS protocol does not have a built-in security mechanism.
Link-level security mechanisms such as MACSec can be employed to prevent eavesdropping on traffic using confidentiality and authentication.
Further, there is a draft for opportunistic encryption for hop-by-hop and end-to-end encryption in MPLS~\cite{ietf-mpls-opportunistic-encrypt-03}.
However, end-to-end encryption is not feasible with hop-by-hop MNA processing as nodes need to inspect the contents of the MPLS stack~\cite{ietf-mpls-mna-fwk-05}.}

\subsection{MNA Information Originating from Other MPLS Domains}
\reviewerii{
An MPLS network must be protected from processing MPLS labels originating outside the network, e.g., label stacks forged by an attacker~\cite{ietf-mpls-mna-fwk-05,ietf-mpls-mna-hdr-04}.
\acp{LSR} must process only label stacks including network actions intended for them.
This is trivial within a single domain, but challenging in a multi-domain context.
We consider the example in \fig{pdfs/ihle31} where traffic from customer domain $D_1$ is tunneled through a provider domain $D_2$ into customer domain $D_3$.
$D_1$ and $D_3$ are under the same administrative control while $D_2$ is not.
LER $R_1$ pushes \acp{LSE} including network actions onto packets that are intended for processing by $R_1$ and $R_4$ but not for processing by $R_2$ and $R_3$.}

\figeps[\columnwidth]{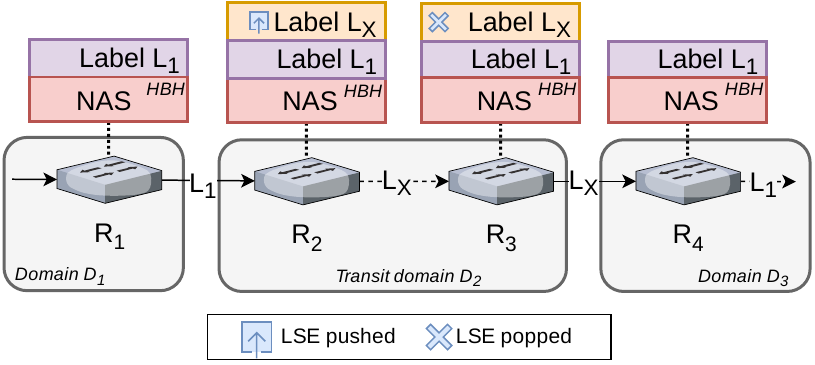}{Traffic between two domains under the same administrative control is tunneled through a provider domain under different administrative control.}

\reviewerii{
In the provider domain $D_2$, the \acp{LER} $R_2$ and $R_3$ must be protected from erroneously processing the \acs{HBH}-scoped \acs{NAS} which is intended only for \acp{LSR} $R_1$ and $R_4$.
Several options are possible.
First, if MNA is not supported in the provider domain, the network actions are not executed and can be forwarded through the provider domain.
However, in this case, the provider domain cannot take advantage of MNA.
Second, if the ingress \acs{LER} has an \acs{RLD} equal to or greater than all other \acp{LSR} in the provider domain, it can drop all incoming packets that contain network actions in reach for any \acs{LSR} of $D_2$.
However, this does not allow MNA traffic to pass through the provider domain.
In a third solution, the ingress \acs{LER} $R_2$ pushes an entire new MPLS stack including a bottom of stack bit so that all \acp{LSR} in the provider domain stop parsing the stack beyond that bit.
This also prevents them from executing the \acs{HBH}-scoped network actions because it isolates the MNA information into two separate domain-specific stacks~\cite{ietf-mpls-mna-fwk-05}.}

\reviewerii{
The \acp{LSR} in the customer domain $D_3$ must be protected from executing network actions that may be inserted by an adversarial provider.
As a simple solution, the ingress \acs{LER} $R_4$ may drop any incoming traffic containing network actions.
However, then network actions cannot be used between $D_1$ and $D_3$ which are under the same administrative control.
To support MNA, the ingress \acs{LER} $R_4$ may forward external packets from a set of non-critical network actions such as NFFRR and block critical network actions such as for network slicing.
As a third solution the customer trusts its service provider and accepts all MNA traffic for execution in domain $D_3$.}

\subsection{Syntax and Semantics of Network Actions}
\reviewerii{
Operators must ensure that ingress LERs enforce syntactically correct \acs{NAS}.
Otherwise, the MPLS stack will be corrupted resulting in packet loss.
In addition, network actions are semantically unbounded, i.e., they are not restricted to a fixed feature set but can describe arbitrary operations.
For example, a network action could be defined that unintentionally enables manipulating a node's memory to compromise forwarding or facilitate eavesdropping.
The IETF must therefore ensure that only secure network actions are defined that do not compromise the security of the network.
Some opcodes are reserved for locally defined network actions.
In this case, implementers and network operators must ensure that these network actions are secure~\cite{ietf-mpls-mna-hdr-04}.
}
\section{Conclusion}
\label{sec:conclusion}
In this paper, we gave an overview of the concepts and the header encoding in the \acs{MNA} framework proposed by the IETF.
We summarized the use cases identified by the working group and gave an outlook on how they can be implemented in the \acs{MNA} framework.
Network actions in the \acs{MNA} framework are conceptually similar to IPv6 extension headers (EH).
However, IPv6 EH are insufficiently supported on hardware platforms, mostly due to their flexible header encoding that allows for large payloads.
Therefore, we verified the feasibility of the encoding introduced with MNA for hardware platforms.

\revieweri{This paper presents P4-MNA, the first implementation of the \acs{MNA} framework.
It shows that the proposed MNA framework is well implementable on programmable hardware such as the Intel Tofino™ 2 switching ASIC with a line speed of \SI{400}{\gbps} per port.
We analyzed the scalability of P4-MNA regarding the number of processable \acp{LSE}, the number of implementable network actions, and the complexity of network actions.
P4-MNA is capable of parsing and processing $51$ \acp{LSE} containing up to $32$ network actions at a line speed of \SI{400}{\gbps}.
The evaluation demonstrated that P4-MNA can efficiently process multiple network actions while maintaining line speed and minimal latency validating its suitability for high-speed networks.
Further, we showed the capability of P4-MNA to implement custom network actions such as link-specific packet loss measurement using \acs{AMM} and bandwidth reservation with network slicing.}
The source code is publicly available on GitHub\cite{p4-mna-git}.

\revieweri{We explained how the \acf{RLD} constrains header stacking in MNA and argued that header stacking is also constrained by maximum \acs{NAS} sizes. 
Therefore, we proposed to make maximum \acs{NAS} sizes a hardware parameter and signal them like the \acs{RLD}.
This facilitates the integration of hardware with smaller \acs{RLD} into MNA-enabled domains.}
A detailed signaling proposal will be introduced in the future.

We explained that some bits of the MPLS header for \acf{ISD} must not be changed by intermediate hops for compatibility reasons so that only \SI{30}{\percent} of the available bits can be used for mutable data.
Therefore, the IETF has also proposed the \acf{PSD} approach where all available bits can be used for mutable data, but \acs{PSD} is considered more complex than \acs{ISD}.
Future work will include an implementation of MNA with \acs{PSD} and its analysis as well as a demonstration of use cases with on-path data collection.

\reviewerii{Finally, we summarized the security risks currently being considered by the working group which include link-level security, MNA information originating from other MPLS domains, and the syntax and semantics of network actions.}

\section*{Erratum}
After publication in IEEE OJCOMS~\cite{IhMe25}, it was identified that Figures \ref{fig:pdfs/ihle11}, \ref{fig:pdfs/ihle12}, and \ref{fig:pdfs/ihle31} were incorrectly rendered due to duplication and formatting issues introduced during production. 
An official Erratum has been published and is linked on the IEEE Xplore page~\cite{erratum}.
The figures included in this version are correct.

\section*{List of Abbreviations}
\begin{acronym}[ECMP] 
    \acro{AD}{ancillary data}
    \acro{AMM}{alternate-marking method}
    \acro{bSPL}{base Special Purpose Label}
    \acro{CBR}{constant bit rate}
    \acro{ECMP}{equal-cost multi-path}
    \acro{EH}{Extension header}
    \acro{eSPL}{extended Special Purpose Label}
    \acro{FRR}{Fast Reroute}
    \acro{HBH}{hop-by-hop}
    \acro{I2E}{ingress-to-egress}
    \acro{ISD}{in-stack data}
    \acro{LER}{Label Edge Router}
    \acro{LSE}{Label Stack Entry}
    \acro{LSP}{Label Switched Path}
    \acro{LSR}{Label Switching Router}
    \acro{MAT}{match+action table}
    \acro{MNA}{MPLS Network Actions}
    \acro{MPLS}{Multiprotocol Label Switching}
    \acro{NAL}{Network Action Length}
    \acro{NASL}{Network Action Sub-stack Length}
    \acro{NAS}{Network Action Sub-stack}
    \acro{NRP}{network resource partition}
    \acro{NSH}{network service header}
    \acro{OAM}{operations, administration and maintenance}
    \acro{P4}{Programming Protocol-independent Packet Processors}
    \acro{PSD}{post-stack data}
    \acro{RLD}{readable label depth}
    \acro{SFC}{service function chaining}
    \acro{SID}{segment identifier}
    \acro{SR}{segment routing}
    \acro{TC}{traffic class}
\end{acronym}

\bibliography{conferences, literature} 

@STRING{ Networks = {{International Telecommunication Network Strategy and Planning Symposium (Networks)}}}

@STRING{ Networking = {{IFIP-TC6 Networking Conference (Networking)}}}

@STRING{ JNCA = {{Journal of Network and Computer Applications (JNCA)}}}

@STRING{ IPv6 = {{ACM International Workshop on IPv6 and the Future of the Internet (IPv6)}}}

@STRING{ Performance = {International Symposium on Computer Performance, Modeling, Measurements and Evaluation (Performance)}}

@STRING{ NetSoft = {{IEEE Conference on Network Softwarization (NetSoft)}}}

@STRING{ Computer   = {{IEEE} Computer Magazine}}

@STRING{ IS = {Information Sciences}}

@STRING{ MNA = {{Mobile Networks \& Applications}}}

@misc{IETF,
    title ="{Internet Engineering Task Force (IETF)}",
    howpublished = {http://www.ietf.org/}
}

@misc{RFC2460,
   author    = {Deering, Stephen and Hinden, Robert},
   title     = {{RFC2460: Internet Protocol Version 6 (IPv6) Specification}},
   year      = 1998,
   month     = dec,
   remark = {ftp://ftp.isi.edu/in-notes/rfc2460.txt}
}

@misc{RFC3031,
   author    = {Rosen, Eric C. and Viswanathan, Arun and Callon, Ross},
   title     = {{RFC3031: Multiprotocol Label Switching Architecture}},
   year      = 2001,
   month     = jan,
   remark = {http://www.ietf.org/rfc/rfc3031.txt}
}

@Misc{RFC4090,
  Author         = "Ping Pan and George Swallow and Alia Atlas",
  Title          = "{RFC4090: Fast Reroute Extensions to RSVP-TE for LSP Tunnels}",
  month          = may,
  year           = 2005,
  remark         = {local repair,{ftp://ftp.rfc-editor.org/in-notes/rfc4090.txt}}
}

@MISC{RFC6790,
  author =       "Clarence {Filsfils, Ed.} and Pierre {Francois, Ed.} and Mike Shand and Bruno Decraene
               and James Uttaro and Nicolai Leyman and Martin Horneffer",
  title =        "{RFC6790: The Use of Entropy Labels in MPLS Forwarding}",
  remark =       "{http://www.rfc-editor.org/rfc/rfc6790.txt}",
  year =         2012,
  month =        nov
}

@INPROCEEDINGS{HeVe17,
  author={Hendriks, Luuk and Velan, Petr and Schmidt, Ricardo de O. and de Boer, Pieter-Tjerk and Pras, Aiko},
  booktitle={Network Traffic Measurement and Analysis Conference (TMA)}, 
  title={{Threats and Surprises behind IPv6 Extension Headers}}, 
  year={2017},
    month = jun,
  pages={1--9},
}

@ARTICLE{NaGa22,
  author={Marlon Naagas and Anazel P Gamilla},
  journal={International Journal of Electrical and Computer Engineering (IJECE)}, 
  title={{Denial of Service Attack: An Analysis to IPv6 Extension Headers Security Nightmares}}, 
  year={2022},
  month = jun,
}

@article{GaNa22,
    author = {Gamilla, Anazel and Naagas, Marlon},
    year = {2022},
    month = feb,
    pages = {319--326},
    title = {{Header of Death: Security Implications of IPv6 Extension Headers to the Open-Source Firewall}},
    volume = {11},
    journal = {Bulletin of Electrical Engineering and Informatics},
}

@misc{rfc8660,
    series =    {Request for Comments},
    number =    8660,
    howpublished =  {RFC 8660},
    publisher = {RFC Editor},
    author =    {Ahmed Bashandy and Clarence Filsfils and Stefano Previdi and Bruno Decraene and Stephane Litkowski and Rob Shakir},
    title =     {{Segment Routing with the MPLS Data Plane}},
    pagetotal = 29,
    year =      2019,
    month =     dec,
}

@misc{rfc9098,
    series =    {Request for Comments},
    number =    9098,
    howpublished =  {RFC 9098},
    publisher = {RFC Editor},
    author =    {Fernando Gont and Nick Hilliard and Gert Döring and Warren "Ace" Kumari and Geoff Huston and Will (Shucheng) LIU},
    title =     {{Operational Implications of IPv6 Packets with Extension Headers}},
    pagetotal = 17,
    year =      2021,
    month =     sep,
}

@article{CuSe24,
    title = {{Is it possible to extend IPv6?}},
    journal = {Computer Communications},
    volume = {214},
    pages = {90--99},
    year = {2024},
    month = jan,
    author = {Ana Custura and Raffaello Secchi and Elizabeth Boswell and Gorry Fairhurst},
}

@techreport{ietf-mpls-mna-hdr-04,
    number =    {draft-ietf-mpls-mna-hdr-12},
    type =      {Internet-Draft},
    institution =   {{IETF}},
    publisher = {{IETF}},
    note =      {Work in Progress},
    url =       {https://datatracker.ietf.org/doc/draft-ietf-mpls-mna-hdr/12/},
    author =    {Jaganbabu Rajamanickam and Rakesh Gandhi and Royi Zigler and Haoyu Song and Kireeti Kompella},
    title =     {{MPLS Network Action (MNA) Sub-Stack Solution}},
    pagetotal = 28,
    year =      2025,
    month =     mar,
    day =       3,
}

@techreport{ietf-mpls-mna-fwk-05,
    number =    {draft-ietf-mpls-mna-fwk-15},
    type =      {Internet-Draft},
    institution =   {{IETF}},
    publisher = {{IETF}},
    note =      {Work in Progress},
    url =       {https://datatracker.ietf.org/doc/draft-ietf-mpls-mna-fwk/15/},
    author =    {Loa Andersson and Stewart Bryant and Matthew Bocci and Tony Li},
    title =     {{MPLS Network Actions (MNA) Framework}},
    pagetotal = 23,
    year =      2024,
    month =     dec,
    day =       27,
}

@techreport{ietf-mpls-ps-mna-hdr-00,
    number =    {draft-ietf-mpls-ps-mna-hdr-00},
    type =      {Internet-Draft},
    institution =   {IETF},
    publisher = {IETF},
    note =      {Work in Progress},
    url =       {https://datatracker.ietf.org/doc/draft-ietf-mpls-ps-mna-hdr/00/},
    author =    {Jaganbabu Rajamanickam and Rakesh Gandhi and Royi Zigler and Tony Li and Jie Dong},
    title =     {{Post-Stack MPLS Network Action (MNA) Solution}},
    pagetotal = 20,
    year =      2025,
    month =     feb,
    day =       16,
}

@techreport{ietf-mpls-mna-usecases-03,
    number =    {{draft-ietf-mpls-mna-usecases-15}},
    type =      {{Internet-Draft}},
    institution =   {{IETF}},
    publisher = {{IETF}},
    note =      {Work in Progress},
    url =       {\url{https://datatracker.ietf.org/doc/draft-ietf-mpls-mna-usecases/15/}},
    author =    {Tarek Saad and Kiran Makhijani and Haoyu Song and Greg Mirsky},
    title =     {{Use Cases for MPLS Network Action Indicators and MPLS Ancillary Data}},
    pagetotal = 14,
    year =      2024,
    month =     oct,
    day =       7,
}

@misc{mpls_wg,
	author = {{MPLS Working Group}},
	title = {{Multiprotocol Label Switching (mpls))}},
	note = "\textit{Last accessed on 09.10.2024}",
	howpublished = {\url{https://datatracker.ietf.org/wg/mpls/documents/}},
}

@misc{rfc9017,
    series =    {Request for Comments},
    number =    9017,
    howpublished =  {{RFC 9017}},
    publisher = {RFC Editor},
    url =       {https://www.rfc-editor.org/info/rfc9017},
    author =    {Loa Andersson and Kireeti Kompella and Adrian Farrel},
    title =     {{Special-Purpose Label Terminology}},
    pagetotal = 8,
    year =      2021,
    month =     apr,
}

@misc{rfc9197,
    series =    {Request for Comments},
    number =    9197,
    howpublished =  {RFC 9197},
    publisher = {RFC Editor},
    author =    {Frank Brockners and Shwetha Bhandari and Tal Mizrahi},
    title =     {{Data Fields for In Situ Operations, Administration, and Maintenance (IOAM)}},
    pagetotal = 40,
    year =      2022,
    month =     may,
}

@misc{rfc9326,
    series =    {Request for Comments},
    number =    9326,
    howpublished =  {RFC 9326},
    publisher = {RFC Editor},
    author =    {Haoyu Song and Barak Gafni and Frank Brockners and Shwetha Bhandari and Tal Mizrahi},
    title =     {{In Situ Operations, Administration, and Maintenance (IOAM) Direct Exporting}},
    pagetotal = 13,
    year =      2022,
    month =     nov,
}

@misc{bmv2,
	author = {{P4 Language Consortium}},
	title = {{GitHub: Behavioural Model Version 2 (BMv2)}},
	note = "\textit{Last accessed on 09.10.2024}",
	howpublished = {\url{https://github.com/p4lang/behavioral-model}},
}

@misc{p4-mna-git,
	author = {{Chair of Communication Networks, University of Tuebingen}},
	title = {{GitHub: P4-MNA}},
	howpublished = {\url{https://github.com/uni-tue-kn/P4-MNA}},
}

@article{MeLi21,
	title = {{Hardware-Based Evaluation of Scalable and Resilient Multicast With BIER in P4}},
	volume = 9,
	journal = {{IEEE Access}},
	author = {Merling, Daniel and Lindner, Steffen and Menth, Michael},
	year = {2021},
	pages = {{34500--34514}},
}

@techreport{mpls-oam,
    number =    {{draft-gandhi-mpls-ioam-12}},
    type =      {{Internet-Draft}},
    institution =   {{IETF}},
    publisher = {{IETF}},
    note =      {Work in Progress},
    url =       {\url{https://www.ietf.org/archive/id/draft-gandhi-mpls-ioam-12.txt}},
    author =    {Rakesh Gandhi and Frank Brockners and Bin Wen and Bruno Decraene and Haoyu Song},
    title =     {{MPLS Network Action for Transporting In Situ OAM Data Fields}},
    year =      2024,
    month =     mar,
    day =       17,
}

@techreport{mb-mpls-ioam-dex-08,
    number =    {{draft-mb-mpls-ioam-dex-08}},
    type =      {Internet-Draft},
    institution =   {{IETF}},
    publisher = {{IETF}},
    note =      {Work in Progress},
    url =       {\url{https://datatracker.ietf.org/doc/draft-mb-mpls-ioam-dex/08/}},
    author =    {Greg Mirsky and Mohamed Boucadair and Tony Li},
    title =     {{Supporting In-Situ Operations, Administration, and Maintenance Direct Export Using MPLS Network Actions}},
    year =      2024,
    month =     jun,
    day =       28,
}

@techreport{song-mpls-extension-header-13,
    number =    {draft-song-mpls-extension-header-13},
    type =      {Internet-Draft},
    institution =   {{IETF}},
    publisher = {{IETF}},
    note =      {Work in Progress},
    url =       {\url{https://datatracker.ietf.org/doc/draft-song-mpls-extension-header/13/}},
    author =    {Haoyu Song and Tianran Zhou and Loa Andersson and Zhaohui (Jeffrey) Zhang and Rakesh Gandhi},
    title =     {{MPLS Network Actions using Post-Stack Extension Headers}},
    year =      2023,
    month =     oct,
    day =       11,
}

@INPROCEEDINGS{HaSt22,
  author={Haeberle, Marco and Steinert, Benjamin and Weiss, Michael and Menth, Michael},
  booktitle={International Conference on Network Softwarization (NetSoft)}, 
  title={{A Caching SFC Proxy Based on eBPF}}, 
  year=2022,
  volume={},
  number={},
  pages={171--179}
}

@misc{rfc8300,
    series =    {Request for Comments},
    number =    8300,
    howpublished =  {RFC 8300},
    publisher = {RFC Editor},
    author =    {Paul Quinn and Uri Elzur and Carlos Pignataro},
    title =     {{Network Service Header (NSH)}},
    pagetotal = 40,
    year =      2018,
    month =     jan,
}

@misc{rfc8402,
    series =    {Request for Comments},
    number =    8402,
    howpublished =  {RFC 8402},
    publisher = {RFC Editor},
    author =    {Clarence Filsfils and Stefano Previdi and Les Ginsberg and Bruno Decraene and Stephane Litkowski and Rob Shakir},
    title =     {{Segment Routing Architecture}},
    pagetotal = 32,
    year =      2018,
    month =     jul,
}

@misc{rfc8595,
    series =    {Request for Comments},
    number =    8595,
    howpublished =  {RFC 8595},
    publisher = {RFC Editor},
    author =    {Adrian Farrel and Stewart Bryant and John Drake},
    title =     {{An MPLS-Based Forwarding Plane for Service Function Chaining}},
    year =      2019,
    month =     jun,
}

@misc{rfc7665,
    series =    {Request for Comments},
    number =    7665,
    howpublished =  {RFC 7665},
    publisher = {RFC Editor},
    author =    {Joel M. Halpern and Carlos Pignataro},
    title =     {{Service Function Chaining (SFC) Architecture}},
    pagetotal = 32,
    year =      2015,
    month =     oct,
}

@misc{rfc5586,
    series =    {Request for Comments},
    number =    5586,
    howpublished =  {RFC 5586},
    publisher = {RFC Editor},
    author =    {Martin Vigoureux and Stewart Bryant and Matthew Bocci},
    title =     {{MPLS Generic Associated Channel}},
    pagetotal = 19,
    year =      2009,
    month =     jun,
}

@misc{rfc9341,
    series =    {Request for Comments},
    number =    9341,
    howpublished =  {RFC 9341},
    publisher = {RFC Editor},
    author =    {Giuseppe Fioccola and Mauro Cociglio and Greg Mirsky and Tal Mizrahi and Tianran Zhou},
    title =     {{Alternate-Marking Method}},
    pagetotal = 22,
    year =      2022,
    month =     dec,
}

@misc{rfc9342,
    series =    {Request for Comments},
    number =    9342,
    howpublished =  {RFC 9342},
    publisher = {RFC Editor},
    author =    {Giuseppe Fioccola and Mauro Cociglio and Amedeo Sapio and Riccardo Sisto and Tianran Zhou},
    title =     {{Clustered Alternate-Marking Method}},
    pagetotal = 24,
    year =      2022,
    month =     dec,
}

@techreport{ietf-teas-ns-ip-mpls-04,
    number =    {{draft-ietf-teas-ns-ip-mpls-05}},
    type =      {Internet-Draft},
    institution =   {{IETF}},
    publisher = {{IETF}},
    note =      {Work in Progress},
    url =       {\url{https://datatracker.ietf.org/doc/draft-ietf-teas-ns-ip-mpls/}},
    author =    {Tarek Saad and Vishnu Pavan Beeram and Jie Dong and Joel M. Halpern and Shaofu Peng},
    title =     {{Realizing Network Slices in IP/MPLS Networks}},
    pagetotal = 32,
    year =      2025,
    month =     mar,
    day =       2,
}

@techreport{cx-mpls-mna-inband-pm-04,
    number =    {draft-cx-mpls-mna-inband-pm-06},
    type =      {Internet-Draft},
    institution =   {{IETF}},
    publisher = {{IETF}},
    note =      {Work in Progress},
    url =       {https://datatracker.ietf.org/doc/draft-cx-mpls-mna-inband-pm/06/},
    author =    {Weiqiang Cheng and Xiao Min and Rakesh Gandhi and Greg Mirsky and Giuseppe Fioccola},
    title =     {{MNA for Performance Measurement with Alternate Marking Method}},
    pagetotal = 6,
    year =      2025,
    month =     mar,
    day =       2,
}

@misc{rfc8372,
    series =    {Request for Comments},
    number =    8372,
    howpublished =  {RFC 8372},
    institution =   {{IETF}},
    publisher = {{IETF}},
    url =       {\url{https://www.rfc-editor.org/info/rfc8372}},
    author =    {Stewart Bryant and Carlos Pignataro and Mach Chen and Zhenbin Li and Greg Mirsky},
    title =     {{MPLS Flow Identification Considerations}},
    pagetotal = 11,
    year =      2018,
    month =     may,
}

@misc{rfc9088,
    series =    {Request for Comments},
    number =    9088,
    howpublished =  {RFC 9088},
    institution =   {{IETF}},
    publisher = {{IETF}},
    url =       {\url{https://www.rfc-editor.org/info/rfc9088}},
    author =    {Xiaohu Xu and Sriganesh Kini and Peter Psenak and Clarence Filsfils and Stephane Litkowski and Matthew Bocci},
    title =     {{Signaling Entropy Label Capability and Entropy Readable Label Depth Using IS-IS}},
    pagetotal = 7,
    year =      2021,
    month =     aug,
}

@misc{rfc9089,
    series =    {Request for Comments},
    number =    9089,
    howpublished =  {RFC 9089},
    publisher = {RFC Editor},
    url =       {\url{https://www.rfc-editor.org/info/rfc9089}},
    author =    {Xiaohu Xu and Sriganesh Kini and Peter Psenak and Clarence Filsfils and Stephane Litkowski and Matthew Bocci},
    title =     {{Signaling Entropy Label Capability and Entropy Readable Label Depth Using OSPF}},
    pagetotal = 8,
    year =      2021,
    month =     aug,
}

@misc{rfc9543,
    series =    {Request for Comments},
    number =    9543,
    howpublished =  {RFC 9543},
    institution =   {{IETF}},
    publisher = {{IETF}},
    url =       {\url{https://www.rfc-editor.org/info/rfc9543}},
    author =    {Adrian Farrel and John Drake and Reza Rokui and Shunsuke Homma and Kiran Makhijani and Luis M. Contreras and Jeff Tantsura},
    title =     {{A Framework for Network Slices in Networks Built from IETF Technologies}},
    pagetotal = 44,
    year =      2024,
    month =     mar,
}

@techreport{ietf-mpls-inband-pm-encapsulation-13,
    number =    {draft-ietf-mpls-inband-pm-encapsulation-18},
    type =      {Internet-Draft},
    institution =   {{IETF}},
    publisher = {{IETF}},
    note =      {Work in Progress},
    url =       {\url{https://datatracker.ietf.org/doc/draft-ietf-mpls-inband-pm-encapsulation/18/}},
    author =    {Weiqiang Cheng and Xiao Min and Tianran Zhou and Jinyou Dai and Yoav Peleg},
    title =     {{Encapsulation For MPLS Performance Measurement with Alternate-Marking Method}},
    pagetotal = 17,
    year =      2024,
    month =     oct,
    day =       3,
}

@techreport{nffrr,
    number =    {{draft-kompella-mpls-nffrr-04}},
    type =      {{Internet-Draft}},
    institution =   {{IETF}},
    publisher = {{IETF}},
    note =      {Work in Progress},
    url =       {\url{https://www.ietf.org/archive/id/draft-kompella-mpls-nffrr-04.html}},
    author =    {Kireeti Kompella and Wen Li},
    title =     {{No Further Fast Reroute}},
    pagetotal = 17,
    year =      2023,
    month =     oct,
    day =       20,
}

@techreport{li-mpls-mna-nffrr-01,
    number =    {draft-li-mpls-mna-nffrr-01},
    type =      {Internet-Draft},
    institution =   {IETF},
    publisher = {IETF},
    note =      {Work in Progress},
    url =       {\url{https://datatracker.ietf.org/doc/draft-li-mpls-mna-nffrr/01/}},
    author =    {Tarek Saad and Israel Meilik and Tony Li and John Drake},
    title =     {{MPLS Network Actions for No Further Fast Reroute}},
    pagetotal = 5,
    year =      2022,
    month =     oct,
    day =       21,
}

@misc{rfc6669,
    series =    {Request for Comments},
    number =    6669,
    howpublished =  {RFC 6669},
    institution =   {IETF},
    publisher = {IETF},
    url =       {\url{https://www.rfc-editor.org/info/rfc6669}},
    author =    {Nurit Sprecher and Luyuan Fang},
    title =     {{An Overview of the Operations, Administration, and Maintenance (OAM) Toolset for MPLS-Based Transport Networks}},
    pagetotal = 21,
    year =      2012,
    month =     jul,
}

@article{kn,
	title = {{A Survey on Data Plane Programming with P4: Fundamentals, Advances, and Applied Research}},
	journal = JNCA,
	volume = 212,
	year = 2023,
    month = mar,
	author = {Frederik Hauser and Marco Häberle and Daniel Merling and Steffen Lindner and Vladimir Gurevich and Florian Zeiger and Reinhard Frank and Michael Menth},
}

@misc{huawei_hash,
	author = {{Huawei}},
	title = {{Configuring the ECMP Load Balancing Mode}},
	note = "\textit{Last accessed on 01.08.2024}",
	howpublished = {\url{https://support.huawei.com/enterprise/en/doc/EDOC1100137942/60585466/configuring-the-ecmp-load-balancing-mode/}},
}

@misc{juniper_hash,
	author = {{Juniper Networks}},
	title = {{Load Balancing MPLS Traffic}},
	note = "\textit{Last accessed on 01.08.2024}",
	howpublished = {\url{https://www.juniper.net/documentation/us/en/software/junos/mpls/topics/topic-map/load-balancing-mpls-traffic.html}},
}

@misc{cisco_hash,
	author = {{Cisco}},
	title = {{ASR9000XR: Load-balancing Architecture and Characteristics}},
	note = "\textit{Last accessed on 05.03.2025}",
	howpublished = {\url{https://community.cisco.com/t5/service-providers-knowledge-base/asr9000-xr-load-balancing-architecture-and-characteristics/ta-p/3124809#field}},
}

@misc{p4,
	author = {{The P4 Language Consortium}},
	title = {{P4 Open Source Programming Language}},
	note = "\textit{Last accessed on 09.10.2024}",
	howpublished = {\url{https://p4.org/}},
}

@misc{p4spec,
	author = {{The P4 Language Consortium}},
	title = {{{$P4_{16}$} Language Specification}},
	month = may,
	year = 2021,
	note = "\textit{Last accessed on 09.10.2024}",
	howpublished = {\url{https://p4.org/p4-spec/docs/P4-16-v1.2.2.pdf}},
}

@misc{tna,
	author = {{Intel{$^{\circledR}$}}},
	title = {{{$P4_{16}$} Intel{$^{\circledR}$} Tofino™ Native Architecture – Public Version}},
	month = apr,
	year = 2021,
	note = "\textit{Last accessed on 09.10.2024}",
	howpublished = {\url{https://github.com/barefootnetworks/Open-Tofino/blob/master/PUBLIC_Tofino-Native-Arch.pdf}},
}

@ARTICLE{p4tg,
author={Lindner, Steffen and Häberle, Marco and Menth, Michael},
journal={{IEEE Access}}, 
title={{P4TG: 1 Tb/s Traffic Generation for Ethernet/IP Networks}}, 
year=2023,
month = feb,
volume=11,
pages={17525--17535}}

@ARTICLE{MaMu21,
  author={Manzanares-Lopez, Pilar and Muñoz-Gea, Juan Pedro and Malgosa-Sanahuja, Josemaria},
  journal={IEEE Access}, 
  title={{Passive In-Band Network Telemetry Systems: The Potential of Programmable Data Plane on Network-Wide Telemetry}}, 
  year={2021},
  volume={9},
  pages={20391--20409}}

@article{FrZa24,
title = {{A Comprehensive Latency Profiling Study of the Tofino P4 Programmable ASIC-based Hardware}},
journal = {Computer Communications},
volume = {218},
pages = {14--30},
year = {2024},
author = {David Franco and Eder {Ollora Zaballa} and Mingyuan Zang and Asier Atutxa and Jorge Sasiain and Aleksander Pruski and Elisa Rojas and Marivi Higuero and Eduardo Jacob},
}

@article{ChYo23,
  title={{Is Large MTU Beneficial to Cellular Core Networks?}},
  author={Young Choi and Jun-Sup Yoon and Younggyoun Moon and KyoungSoo Park},
  journal={Asia-Pacific Workshop on Networking},
  year={2023},
  month=jun
}

@Article{LiLv22,
AUTHOR = {Liu, Zhongpei and Lv, Gaofeng and Wang, Jichang and Yang, Xiangrui},
TITLE = {{A Programmable SRv6 Processor for SFC}},
JOURNAL = {MDPI Electronics},
VOLUME = {11},
YEAR = {2022},
NUMBER = {18},
month = sep
}

@inproceedings{LeBo17,
author = {Lebrun, David and Bonaventure, Olivier},
title = {{Implementing IPv6 Segment Routing in the Linux Kernel}},
year = {2017},
booktitle = {Applied Networking Research Workshop},
pages = {35–41},
numpages = {7},
month = jul
}

@misc{Ta19,
	author = {{Chuck Tato}},
	title = {{Segment Routing Over IPv6 Acceleration Using Intel FPGA Programmable Acceleration Card N3000}},
	month = oct,
	year = 2019,
	note = "\textit{Last accessed on 17.12.2024}",
	howpublished = {https://www.intel.com/content/dam/www/central-libraries/us/en/documents/wp-01295-hcl-segment-routing-over-ipv6-acceleration-using-intel-fpga-programmable-acceleration-card-n3000-white-paper.pdf},
}

@inproceedings{IuDo20,
        AUTHOR = {Iurman, Justin and Donnet, Benoît and Brockners, Frank},
        TITLE = {{Implementation of IPv6 IOAM in Linux Kernel}},
        YEAR = {2020},
        month = aug,
        BOOKTITLE = {Netdev},
}

@ARTICLE{ChCh22,
  author={Chen, Yan-Wei and Li, Chi-Yu and Tseng, Chien-Chao and Hu, Min-Zhi},
  journal={IEEE Transactions on Network and Service Management}, 
  title={{P4-TINS: P4-Driven Traffic Isolation for Network Slicing With Bandwidth Guarantee and Management}}, 
  year={2022},
  volume={19},
  number={3},
  pages={3290--3303},
}

@techreport{li-mpls-mna-nrp-selector-01,
    number =    {draft-li-mpls-mna-nrp-selector-01},
    type =      {Internet-Draft},
    institution =   {IETF},
    note =      {Work in Progress},
    url =       {\url{https://datatracker.ietf.org/doc/draft-li-mpls-mna-nrp-selector/01/}},
    author =    {Tony Li and John Drake and Vishnu Pavan Beeram and Tarek Saad and Israel Meilik},
    title =     {{MPLS Network Actions for Network Resource Partition Selector}},
    pagetotal = 8,
    year =      2024,
    month =     jun,
    day =       25,
}

@inproceedings{IhZi24,
      title={{Enhancements to P4TG: Protocols, Performance, and Automation}},
      author={Fabian Ihle and Etienne Zink and Steffen Lindner and Michael Menth},
      year={2025},
      booktitle={KuVS Workshop on Network Softwarization (KuVS NetSoft)},
      month=apr,
}

@inproceedings{ZiFl25,
      title={{Rust Barefoot Runtime (RBFRT): Fast Runtime Control for the Intel Tofino}},
      author={Etienne Zink and Moritz Flüchter and Steffen Lindner and Fabian Ihle and Michael Menth},
      year={2025},
      booktitle={KuVS Workshop on Network Softwarization (KuVS NetSoft)},
      month=apr,
}

@inproceedings{LeIu22,
author = {L\'{e}as, Rapha\"{e}l and Iurman, Justin and Vyncke, \'{E}ric and Donnet, Benoit},
title = {{Measuring IPv6 Extension Headers Survivability with James}},
year = {2022},
booktitle = {{ACM Internet Measurement Conference}},
pages = {746–747},
}

@techreport{ietf-mpls-opportunistic-encrypt-03,
    number =    {draft-ietf-mpls-opportunistic-encrypt-03},
    type =      {Internet-Draft},
    institution =   {IETF},
    publisher = {IETF},
    note =      {Work in Progress},
    url =       {https://datatracker.ietf.org/doc/draft-ietf-mpls-opportunistic-encrypt/03/},
    author =    {Adrian Farrel and Stephen Farrell},
    title =     {{Opportunistic Security in MPLS Networks}},
    pagetotal = 38,
    year =      2017,
    month =     mar,
    day =       28,
}

@ARTICLE{IhMe25,
  author={Ihle, Fabian and Menth, Michael},
  journal={IEEE Open Journal of the Communications Society}, 
  title={{MPLS Network Actions: Technological Overview and P4-Based Implementation on a High-Speed Switching ASIC}}, 
  year={2025},
month=apr,
  volume={6},
  number={},
  pages={3480--3501}}

@ARTICLE{erratum,
  author={Ihle, Fabian and Menth, Michael},
  journal={IEEE Open Journal of the Communications Society}, 
  title={{Erratum to “MPLS Network Actions: Technological Overview and P4-Based Implementation on a High-Speed Switching ASIC”}}, 
  year={2025},
  volume={6},
  number={},
  pages={4341--4341},
}
\bibliographystyle{ieeetr}

\begin{IEEEbiography}
	[{\includegraphics[width=1in,height=1.25in,clip,keepaspectratio] 
		{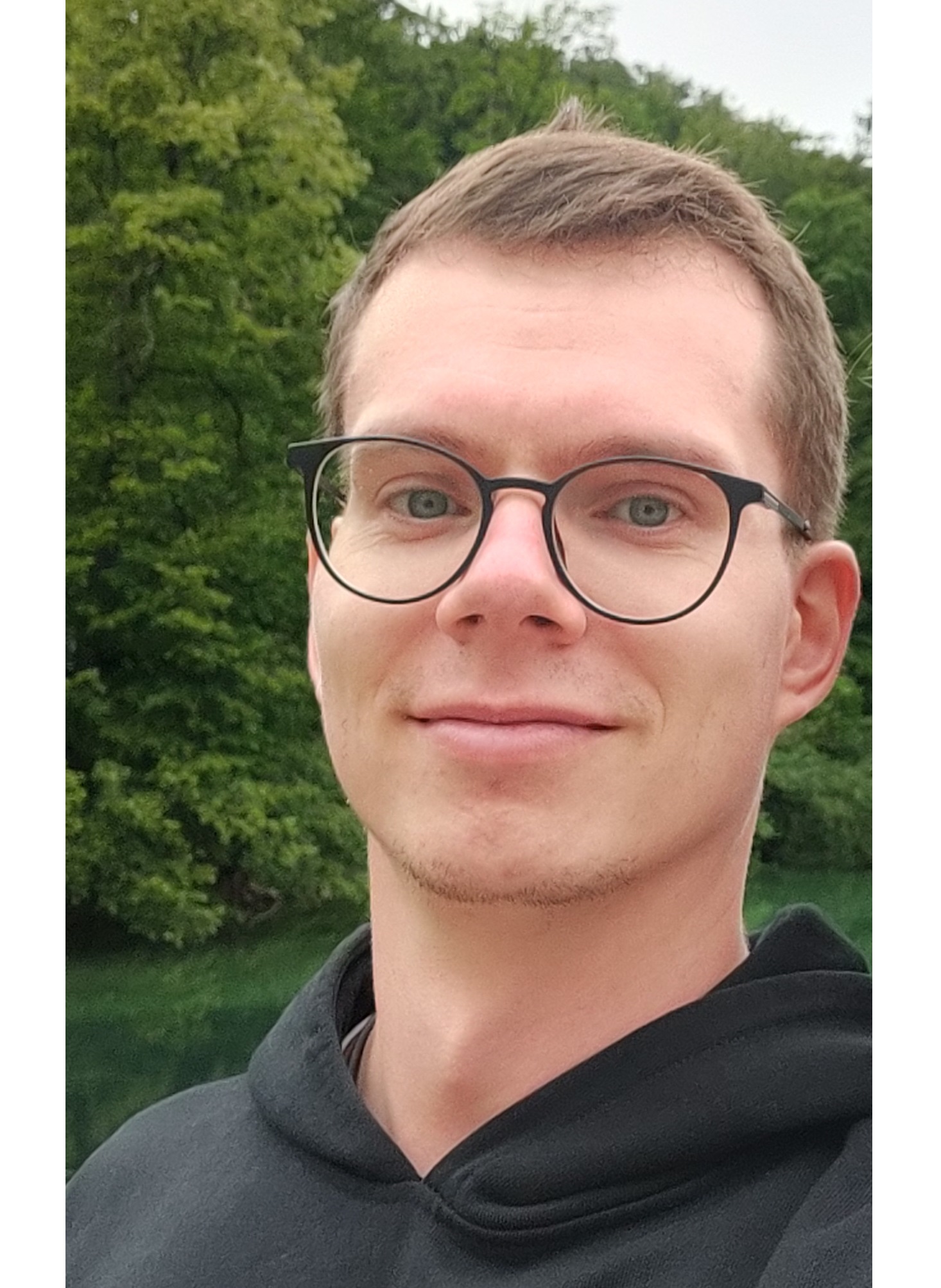}}]{Fabian Ihle }
    received his bachelor's (2021) and master's degrees (2023) in computer science at the University of Tuebingen. Afterwards, he joined the communication networks research group of Prof. Dr. habil. Michael Menth as a Ph.D. student.
    His research interests include software-defined networking, P4-based data plane programming, resilience, and Time-Sensitive Networking (TSN).
\end{IEEEbiography}

\begin{IEEEbiography}
	[{\includegraphics[width=1in,height=1.25in,clip,keepaspectratio]
		{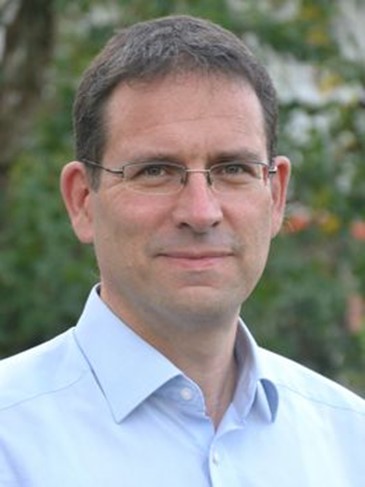}}]{Michael Menth, }
	(Senior Member, IEEE) is professor at the Department
of Computer Science at the University of Tuebingen/Germany and
chairholder of Communication Networks since 2010. He studied,
worked, and obtained diploma (1998), PhD (2004), and habilitation
(2010) degrees at the universities of Austin/Texas, Ulm/Germany,
and Wuerzburg/Germany. His special interests are performance
analysis and optimization of communication networks, resilience and
routing issues, as well as resource and congestion management. His
recent research focus is on network softwarization, in particular
P4-based data plane programming, Time-Sensitive Networking (TSN),
Internet of Things, and Internet protocols. Dr. Menth contributes
to standardization bodies, notably to the IETF.
\end{IEEEbiography}

\end{document}